\documentclass[useAMS,fleqn,usenatbib]{mnras}

\usepackage{graphicx}

\usepackage[T1]{fontenc}
\usepackage{ae,aecompl}
\usepackage{hyperref}

\newcommand{\ncomp}{N_{\!\mbox{\scriptsize\it comp}}}
\newcommand{\sigmaobs}{\sigma_{\!\mbox{\scriptsize\it obs}}}
\newcommand{\sigmainstr}{\sigma_{\!\mbox{\scriptsize\it instr}}}
\newcommand{\sigmath}{\sigma_{\!\mbox{\scriptsize\it th}}}
\newcommand{\sigmafs}{\sigma_{\!\mbox{\scriptsize\it fs}}}

\newcommand{\reffmath}{R_{\!\mbox{\scriptsize\it eff}}}

\newcommand{\msun}{$M_\odot$}

\newcommand{\hi}{H\,{\sc i}\rm}
\newcommand{\hii}{H\,{\sc ii}\rm}

\newcommand{\hei}{He\,{\sc i}\rm}

\newcommand{\oiii}{[O\,{\sc iii}]}

\newcommand{\sii}{[S\,{\sc ii}]}

\newcommand{\eg}{{e.g.}}
\newcommand{\ie}{{i.e.}}

\newcommand{\te}{$T_e$}

\newcommand{\lin}{$\,\lambda$}

\newcommand{\oh}{12\,+\,log(O/H)}

\newcommand{\vs}{vs.}

\newcommand{\halpha}{H$\alpha$}
\newcommand{\hbeta}{H$\beta$}

\newcommand{\kms}{\,km\,s$^{-1}$}

\title[Internal kinematics of giant H\,{\normalsize \textit{II}} regions in M101]{Internal kinematics of giant H\,{\Large \textbf{II}} regions in M101 with the Keck Cosmic Web Imager}
\author[F.~Bresolin et al.]{Fabio~Bresolin$^{1}$\thanks{E-mail:
bresolin@ifa.hawaii.edu}, Luca Rizzi$^{2}$, I-Ting Ho$^{3}$, Roberto Terlevich$^{4,5}$,
\newauthor Elena Terlevich$^{4}$, Eduardo Telles$^{6}$, Ricardo Ch\'avez$^{7}$,
\newauthor Spyros Basilakos$^{8,9}$ and Manolis Plionis$^{8,10}$
\\ \\
$^{1}$Institute for Astronomy, 2680 Woodlawn Drive, Honolulu, HI 96822, USA\\
$^{2}$W.M. Keck Observatory, 65-1120 Mamalahoa Highway, Kamuela, HI 96743, USA \\
$^{3}$Max Planck Institut f\"ur Astronomie, K\"onigstuhl 17, 69117 Heidelberg, Germany \\
$^{4}$Instituto Nacional de Astrofi\'isica \'Optica y Electr\'onica, AP 51 y 216, 72000, Puebla, M\'exico \\
$^{5}$Institute of Astronomy, University of Cambridge, Madingley Road, Cambridge CB3 0HA, UK \\
$^{6}$Observatorio Nacional, Rua Jos\'e Cristino 77, 20921-400 Rio de Janeiro, Brasil \\
$^{7}$CONACYT-Instituto de Radioastronom\'ia y Astrof\'isica, UNAM, Campus Morelia, C.P. 58089, Morelia, M\'exico\\
$^{8}$National Observatory of Athens, P.Pendeli 15236, Athens, Greece\\
$^{9}$Physics Department, Aristotle University of Thessaloniki, Thessaloniki 54124, Greece\\
$^{10}$Academy of Athens Research Center for Astronomy \& Applied Mathematics, Soranou Efessiou 4, Athens 11-527, Greece
} 

\pubyear{2020}

\begin{document}
\label{firstpage}
\pagerange{\pageref{firstpage}--\pageref{lastpage}}
\maketitle


\begin{abstract}		
\noindent
We study the kinematics of the giant \hii\ regions NGC~5455 and NGC~5471 located in the galaxy M101, using 
integral field observations that include the \hbeta\ and \oiii\lin5007 emission lines, obtained with the Keck Cosmic Web Imager.
We analyse the line profiles using both single and multiple Gaussian curves, gathering evidence for the presence of several expanding shells and moving filaments. 
The line decomposition shows that a broad ($\sigma\simeq 30-50$\,\kms) underlying component is ubiquitous, extending across hundreds of pc, while a large fraction of the narrow components have subsonic line widths.
The supersonic turbulence inferred from the global line profiles is consistent with the velocity dispersion of the individual narrow components, \ie\ the global profiles likely arise from the combined contribution of discrete gas clouds.  We confirm the presence of very extended (400\,--\,1200\,\kms) low-intensity line components in three bright star-forming cores in NGC~5471, possibly representing kinematic signatures of supernova remnants. For one of these, the known supernova remnant host NGC~5471\,B, 
we find a significantly reduced \oiii/\hbeta\ line ratio relative to the surrounding photoionized gas, due to the presence of a radiative shock at low metallicity.
We explore the systematic width discrepancy between \hi\ and \oiii\ lines, present in both global and individual spaxel spectra. 
We argue that the resolution of this long-standing problem lies in the physics of the line-emitting gas rather than in the smearing effects induced by the different thermal widths.

\end{abstract}


\section{Introduction}

The supersonic stirring of the gas in giant \hii\ regions, first inferred from the
integrated line widths of the most luminous ionized nebulae in the galaxies M33 and M101 by \citet{Smith:1970}, remains a  poorly explored and understood phenomenon. 
Some of the mechanisms put forward to interpret the presence of these supersonic motions (including hydrodynamic turbulence and champagne flows) have not survived deeper scrutiny
(\citealt{Melnick:1987, Yang:1996}). However, a dispute has persisted in the literature concerning the role of two alternative processes, namely self gravity (\citealt{Terlevich:1981, Tenorio-Tagle:1993}) and the energetics of massive stars (\citealt{Gallagher:1983, Chu:1994}). 
Our motivation for a new work on this subject is that a better understanding of supersonic motions in giant \hii\ regions is crucial in order to use these objects as extragalactic distance indicators via the relationship between their integrated emission line widths and luminosities
(\citealt{Chavez:2014, Fernandez-Arenas:2018, Gonzalez-Moran:2019}).

Through spatially-resolved spectroscopy of the supergiant \hii\ region NGC~604 in M33, \citet{Yang:1996} and \citet{Munoz-Tunon:1996} reached the conclusion that, once corrected for thermal motions, the supersonic line widths can be accounted for, in approximately equal proportions, by virial motions and stellar winds. The latter produce fast-moving gas filaments and expanding shells, leading to the broad wings observed in the integrated line profiles.

In our vicinity, 30 Doradus (30 Dor) in the Large Magellanic Cloud is the quintessential giant \hii\ region. Its integrated line profile stems from the combined contribution of discrete, macroscopic components (shells, filaments and clouds), characterized by subsonic line widths, 
as shown by \citet{Chu:1994} and \citet{Melnick:1999}. \citet{Melnick:2019} have recently argued that the motion of these discrete components inside the gravitational potential of the whole star forming region originates the supersonic width of the integrated line core, while stellar winds are responsible for the extended line wings.

Testing a paradigm that accounts for the line profiles of giant \hii\ regions by invoking both gravity and stellar winds is a difficult task when looking, by necessity, beyond 30~Dor, because of the dramatic loss in spatial resolution.
We can partly compensate by targeting similar or, perhaps more interestingly, more luminous nebulae
using two-dimensional spectroscopy that extends the kinematics coverage over larger areas than currently feasible in 30~Dor ($0.15 \times 0.15$~kpc$^2$, \citealt{Torres-Flores:2013}).
In this paper we focus on the internal gas kinematics of two of the objects included in the work by
\citet{Smith:1970}, NGC~5455 and NGC~5471. Both are significantly more luminous than 30~Dor (\citealt{Kennicutt:1984}). 

Both NGC~5455 and NGC~5471 have been the focus of previous kinematic studies, carried out utilizing echelle spectrographs and Fabry-Perot interferometers 
(\eg\ \citealt{Smith:1970, Melnick:1977, Hippelein:1986}). 
\cite{Skillman:1984} discovered that the asymmetric line profiles they observed in the bright cores of the \hii\ regions in their sample, which included NGC~5455 and NGC~5471, required a decomposition into both a broad and a narrow Gaussian components.
Line splittings and asymmetric line profiles in giant \hii\ regions were confirmed by subsequent studies (\eg\ \citealt{Chu:1994, Sabalisck:1995}). Other authors focused on high-velocity gas motions and known or suspected supernova remnants (SNRs) in NGC~5471 (\citealt[=\,CK86]{Chu:1986}; \citealt{Castaneda:1990, Chen:2002}), finding 
evidence for broad profiles underlying brighter, narrow components.

The work by \citet{Munoz-Tunon:1995} was among the pioneering efforts made to obtain two-dimensional spectroscopy of extragalactic \hii\ regions, using a Fabry-Perot imaging spectrograph in order to follow the \halpha\ line profile across 
a $1.1\times1.1$~arcmin$^2$ field in NGC~604, NGC~5461 and NGC~5471. These authors detected split line profiles in low surface brightness portions of NGC~5471, and measured supersonic gas motions throughout the extent of this object. Broader lines were measured for two specific knots of star formation, one of which displayed a clearly non-Gaussian line profile.
Our work expands on this study using improved sensitivity and spectral resolution, 
confirming some of the conclusions by \citet{Munoz-Tunon:1995}, but also finding significant deviations from their study once a multi-component line analysis is implemented.

Our paper is organized as follows: we present our integral field data in Sect.~2, and global gas and line profiles in Sec.~3. We analyse the line profiles at each spaxel using single Gaussian fits and multiple Gaussian components in Sec.~4 and 5, respectively. In Sect.~6 we interpret the global supersonic line profiles, based on the results obtained from the multiple Gaussian fits. Sect.~7 focuses on the supernova remnant in NGC~5471\,B. We consider the long-standing problem of the line width discrepancy between recombination and collisionally excited lines in Sect.~8. We conclude in Sect.~9 with a summary of our results.

\section{Observations and data reduction}\label{sec:observations}
We observed the giant \hii\ regions NGC~5455 and NGC~5471 in the galaxy M101 with the Keck Cosmic Web Imager (KCWI) integral
field spectrograph (\citealt{Morrissey:2018}) on the night of Feb 9, 2018. With the selected slicer configuration (0\farcs35 slice width) the field of view of the instrument is $8.4 \times 20.4$ arcsec$^2$. Three overlapping tiles observed 
contiguously yielded a total coverage of $21.4 \times 20.4$ arcsec$^2$ with a scale of 0.146 arcsec\,pixel$^{-1}$. 

With the BH3 grating utilized for the observations the resulting spectra have a FWHM resolution of 0.30~\AA, as measured from arc calibration frames, corresponding to a FWHM velocity resolution
of 18.0~km\,s$^{-1}$ at 5007~\AA\ ($R \simeq 16,700$). The Gaussian instrumental broadening is therefore $\sigmainstr = 7.6$~km\,s$^{-1}$.
A grating central wavelength of 4900~\AA\ allowed us to fit simultaneously the \hbeta, \oiii\lin4959 and \oiii\lin5007 nebular lines
within the wavelength range covered on the detector (4585\,--\,5195~\AA). 

We obtained sequences of 300\,s and 900\,s exposures for NGC~5455, and 180\,s for the brighter NGC~5471. Since the image quality varied substantially between different frames, we decided to retain for our analysis only the best-seeing ($\sim$1 
arcsec\footnote{1~arcsec = 33~pc at an adopted distance of 6.8~Mpc (\citealt{Fernandez-Arenas:2018}).}) data.
Consequently, the total exposure time per single tile amounts to 900\,s in the case of NGC~5455 and 360\,s in the case of NGC~5471. Despite the short integrations, the \oiii\lin5007 line saturates in a few pixels of the brightest region of NGC~5471. In our line profile analysis we therefore used the information derived from the \oiii\lin4959 line for these pixels.

\begin{figure}
	\center
	\includegraphics[width=0.95\columnwidth]{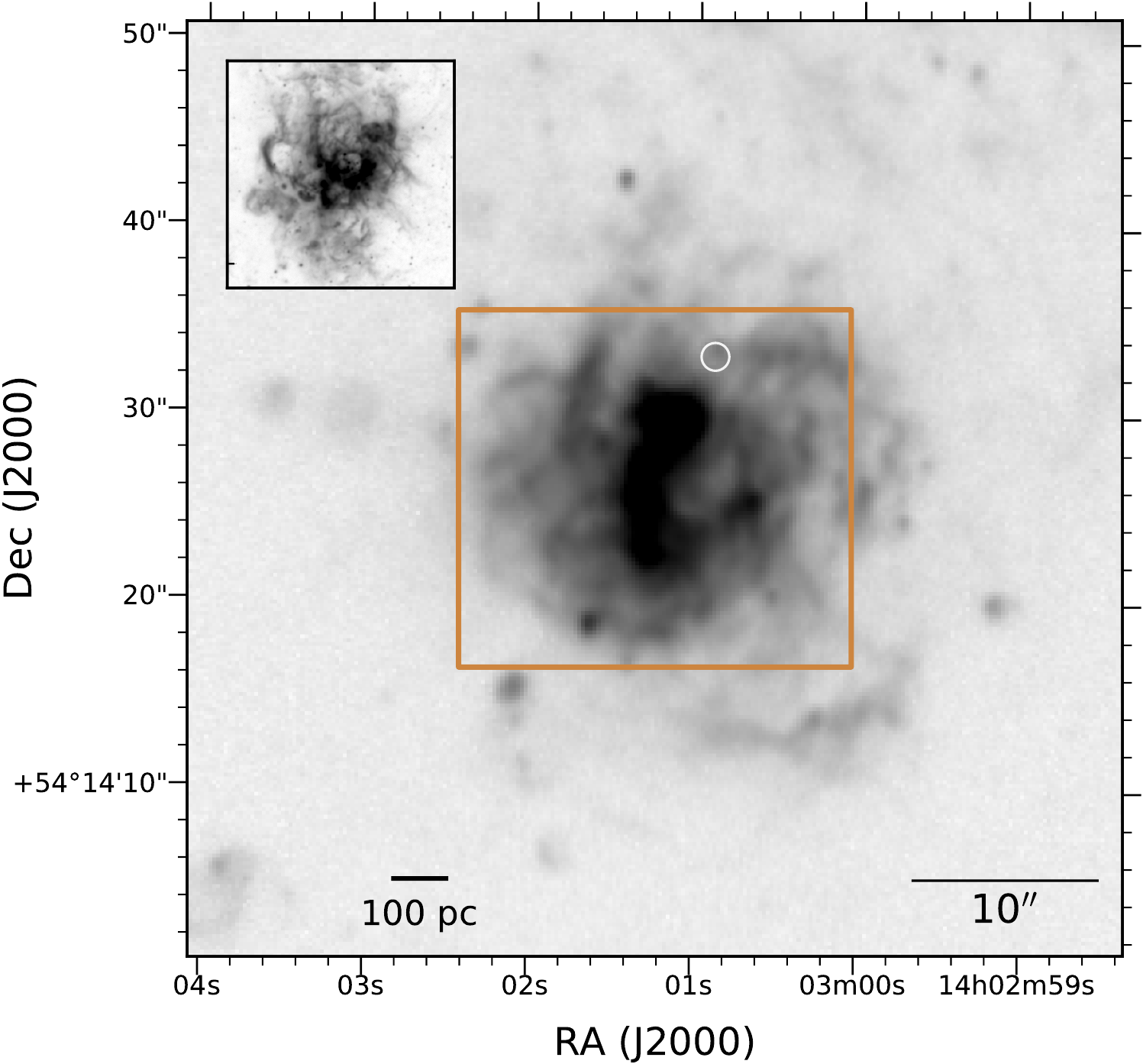}
	\caption{\halpha\ image of NGC~5455 (extracted from the CFHT Science Archive) with overlay of the field observed with KCWI (rectangle). In the inset at the upper left the giant \hii\ region NGC~604 in the galaxy M33 is shown at the same physical scale, for comparison. The circle indicates the position of SN\,1970G
	(\citealt{Dittmann:2014}) for reference. 
		\label{n5455ha}}
\end{figure}

\begin{figure}
	\center
	\includegraphics[width=0.95\columnwidth]{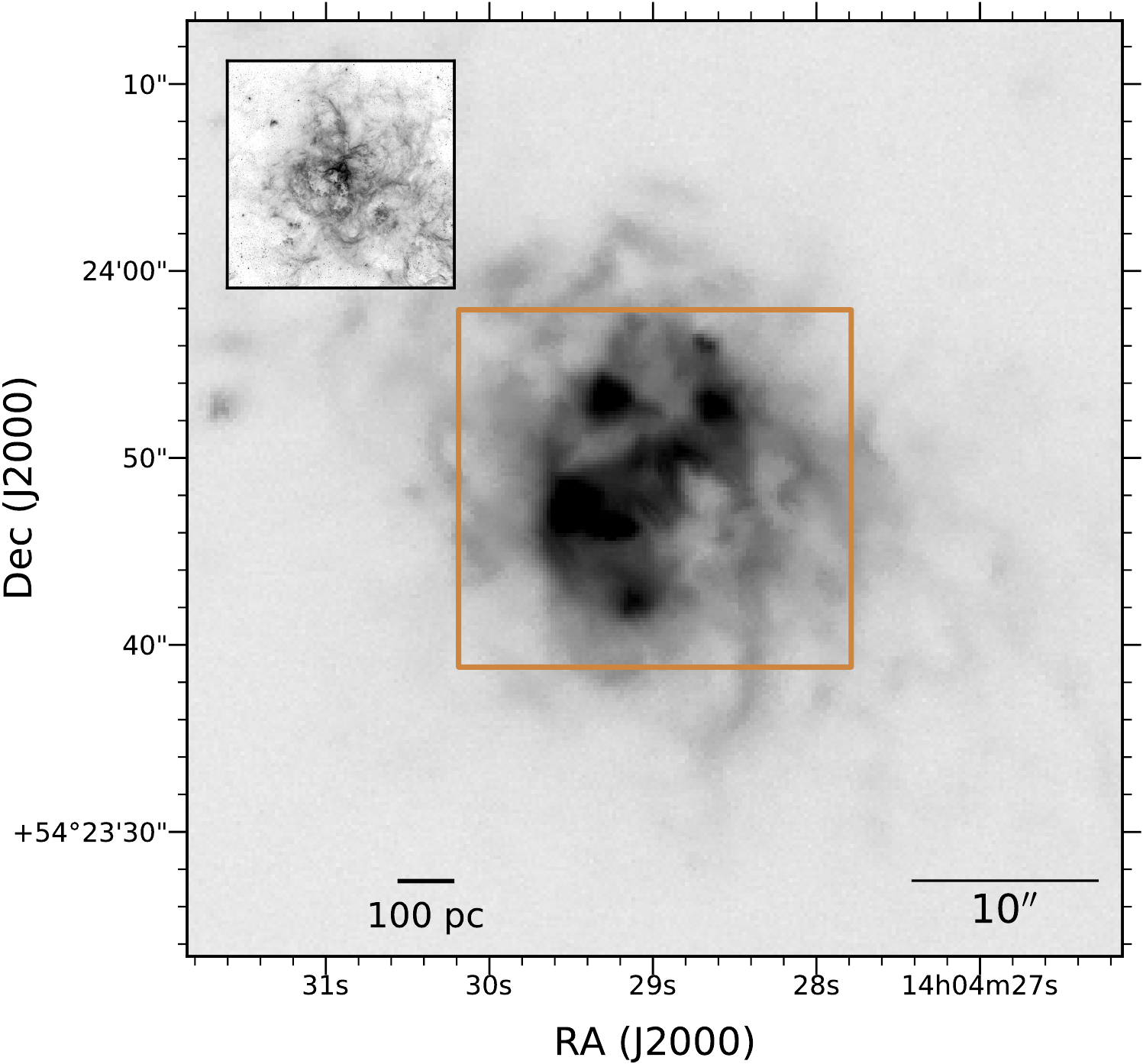}
	\caption{\halpha\ image of NGC~5471 (extracted from the CFHT Science Archive) with overlay of the field observed with KCWI (rectangle). In the inset at the upper left the giant \hii\ region 30~Dor in the LMC is shown at the same physical scale, for comparison (image from the ESO archive).
		\label{n5471ha}}
\end{figure}

Using the data reduction pipeline (described by \citealt{Morrissey:2018}) we produced geometrically corrected, wavelength calibrated
data cubes that account for the effects of differential atmospheric refraction. We then mosaicked and combined the three tiles obtained for each \hii\ region with the {\sc dpuser}\footnote{\url{http://www.mpe.mpg.de/~ott/dpuser}.} software 
(\citealt{Ott:2012a})
and {\sc iraf}\footnote{{\sc iraf} is distributed by the National Optical Astronomy Observatories, which are operated by the Association of Universities for Research in Astronomy, Inc., under cooperative agreement with the National Science Foundation.} routines. In order to facilitate the subsequent analysis
we spatially rebinned the images by a factor of three (1 pixel = 0.44 arcsec). Fig.~\ref{n5455ha} and ~\ref{n5471ha} 
display narrow-band \halpha\ images of NGC~5455 and NGC~5471, respectively, illustrating the spatial extent of the data cubes. 
It can be seen that the KWCI data fully cover the bright portion of the star forming region in both cases, 
extending over $\sim$$0.69 \times 0.63$~kpc$^{2}$ with a spatial resolution of $\sim$30~pc, and excluding only relatively faint outlying gaseous filaments.

\section{Global gas properties and line profiles}\label{Sec:global}
\begin{figure}
	\center
	\includegraphics[width=\columnwidth]{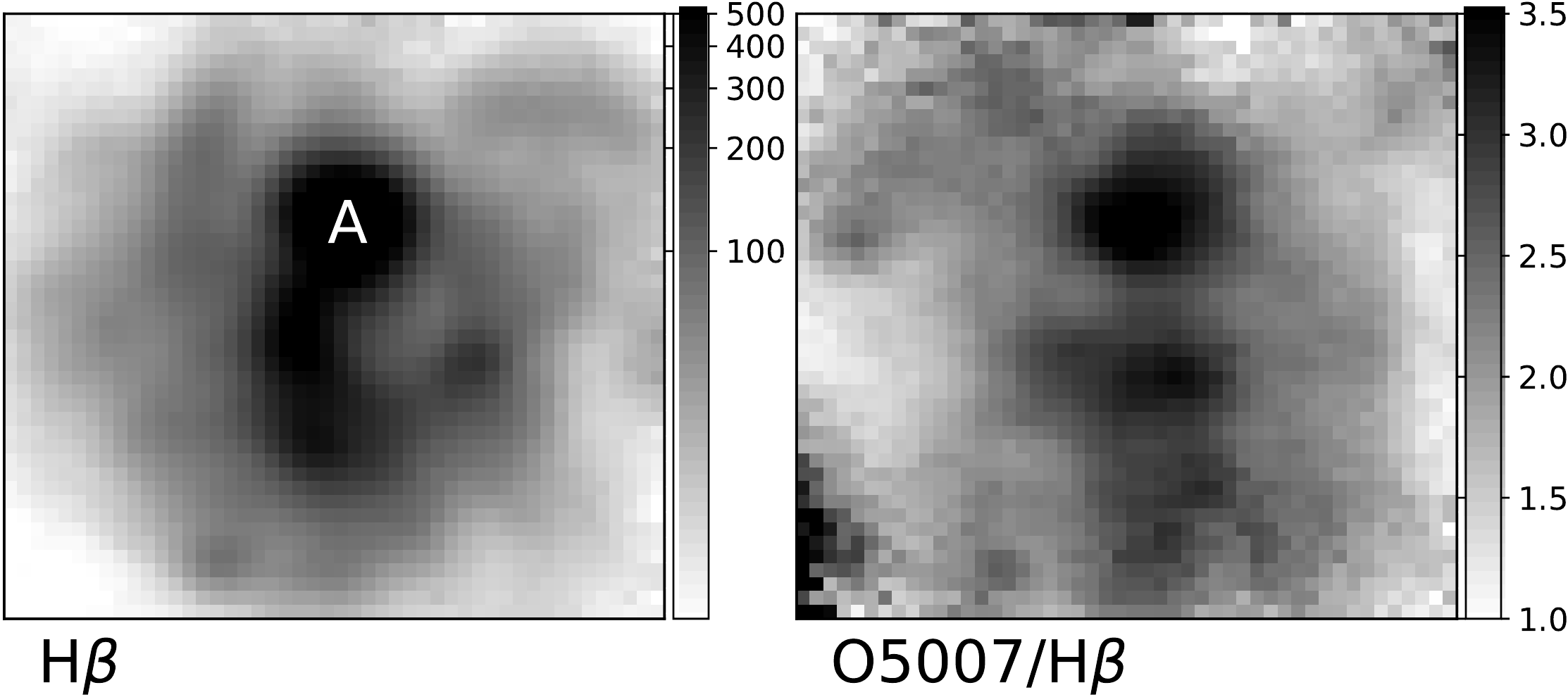}
	\caption{(Left) Continuum-subtracted \hbeta\ image of NGC~5455 derived from the KCWI data cube. The spot with the brightest emission is labelled as A. (Right) The \oiii\lin 5007/\hbeta\ emission line ratio image. In this and subsequent figures we refer to 
		\oiii\lin5007 as O5007.
		\label{n5455nb}}
\end{figure}

\begin{figure}
	\center
	\includegraphics[width=\columnwidth]{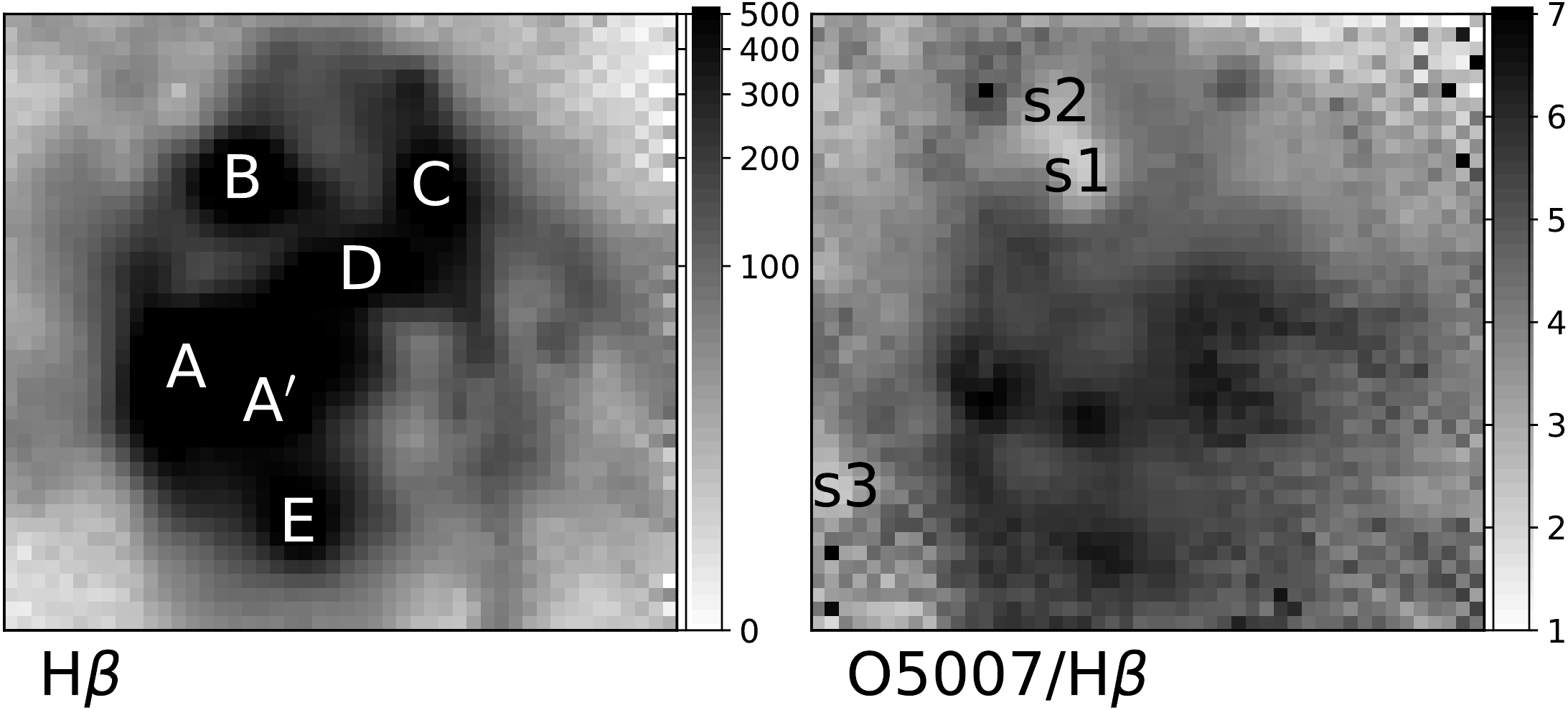}
	\caption{(Left) Continuum-subtracted \hbeta\ image of NGC~5471 derived from the KCWI data cube. The letters A\,--\,E refer to the emission peaks identified by \citet{Skillman:1985}, while A$^\prime$ follows from \citet{Munoz-Tunon:1995}.
		(Right) The \oiii\lin 5007/\hbeta\ emission line ratio image. The three shells identified by \citet{Chen:2002} are labelled as s1, s2 and s3. 
		\label{n5471nb}}
\end{figure}

The KCWI data cubes were handled with the {\tt spectral-cube} package, which is part of the Astropy Project
(\citealt{Astropy-Collaboration:2013, Astropy-Collaboration:2018}). As a first step, in order to visualize 
the spatial distribution of the gas emission properties in the two \hii\ regions, 
we produced continuum-subtracted narrow-band images corresponding to the \hbeta\ and \oiii\ line emission. In Fig.~\ref{n5455nb} (for NGC~5455) and \ref{n5471nb} (for NGC~5471) we show the \hbeta\ line emission map (left panel) and the \oiii\lin5007/\hbeta\ line ratio map (right panel) thus obtained. In these images, as in all other images and maps presented in this paper, North is at the top and East to the left.

In the case of NGC~5455 (Fig.~\ref{n5455nb}) we have labelled as `A' the single, bright blob in the northern section of the nebula. Although this is not obvious because of the stretch used to display the map, region A dominates the overall line emission of NGC~5455. The gas excitation, measured by the \oiii/\hbeta\ line ratio (right panel of Fig.~\ref{n5455nb}), peaks at the position of blob A, but its spatial distribution 
does not fully correlate with that of the \hbeta\ flux in the remaining parts of the nebula.

NGC~5471 is a giant \hii\ region containing multiple ionizing clusters.
In the left panel of Fig.~\ref{n5471nb} we have identified the five positions (A through E) defined by \citet{Skillman:1985}, where 
A corresponds to the brightest star-forming knot in the region -- comparable in luminosity and ionizing cluster mass to 30 Dor (\citealt{Chen:2005a}) -- while B hosts a SNR, identified from optical excitation and nonthermal radio emission properties (\citealt{Skillman:1985}) and characterized by broad emission lines (CK86, \citealt{Chen:2002} -- the latter authors favour the notion of a hypernova remnant). The labeling of the object immediately to the west of A as A$^\prime$ follows \citet{Munoz-Tunon:1995}.

While the gas excitation is high throughout most of the extent of NGC~5471, and in general where the \hbeta\ emission peaks, as seen in the right panel of Fig.~\ref{n5471nb}, there are regions of localized lower excitation, corresponding with the three shells identified by \citet{Chen:2002} on the basis of enhanced \sii/\halpha\ line emission, the classic optical signature for the presence of a radiative shock (\citealt{Mathewson:1973}).
One of these (`s1' in Fig.~\ref{n5471nb}) is associated with knot B and its SNR. \citet{Chen:2002} suggested that also the remaining two shells are related to the presence of SNRs. 
\oiii-to-Balmer line ratios that are low in comparison to nearby photoionized gas have been presented for SNRs before (see \citealt{Shull:1983} for objects in the LMC).
The explanation for the significant drop in the \oiii\lin5007/\hbeta\ line ratio seen in Fig.~\ref{n5471nb} in correspondence of shocked gas regions lies in the low metallicity of NGC~5471, \oh~$\simeq$~8.05 (\citealt{Kennicutt:2003}), equalling that of the Small Magellanic Cloud. At this metallicity the \oiii\lin5007/\hbeta\ ratio predicted by shock models can be significantly lower than for photoionized gas (\eg\ compare Fig.~21 of \citealt{Allen:2008} with the \hii\ region diagnostic diagram shown in Fig.~4 by \citealt{Bresolin:2012}).

\begin{figure*}
	\centering
	\begin{minipage}[t]{.45\textwidth}
		\includegraphics[width=\columnwidth]{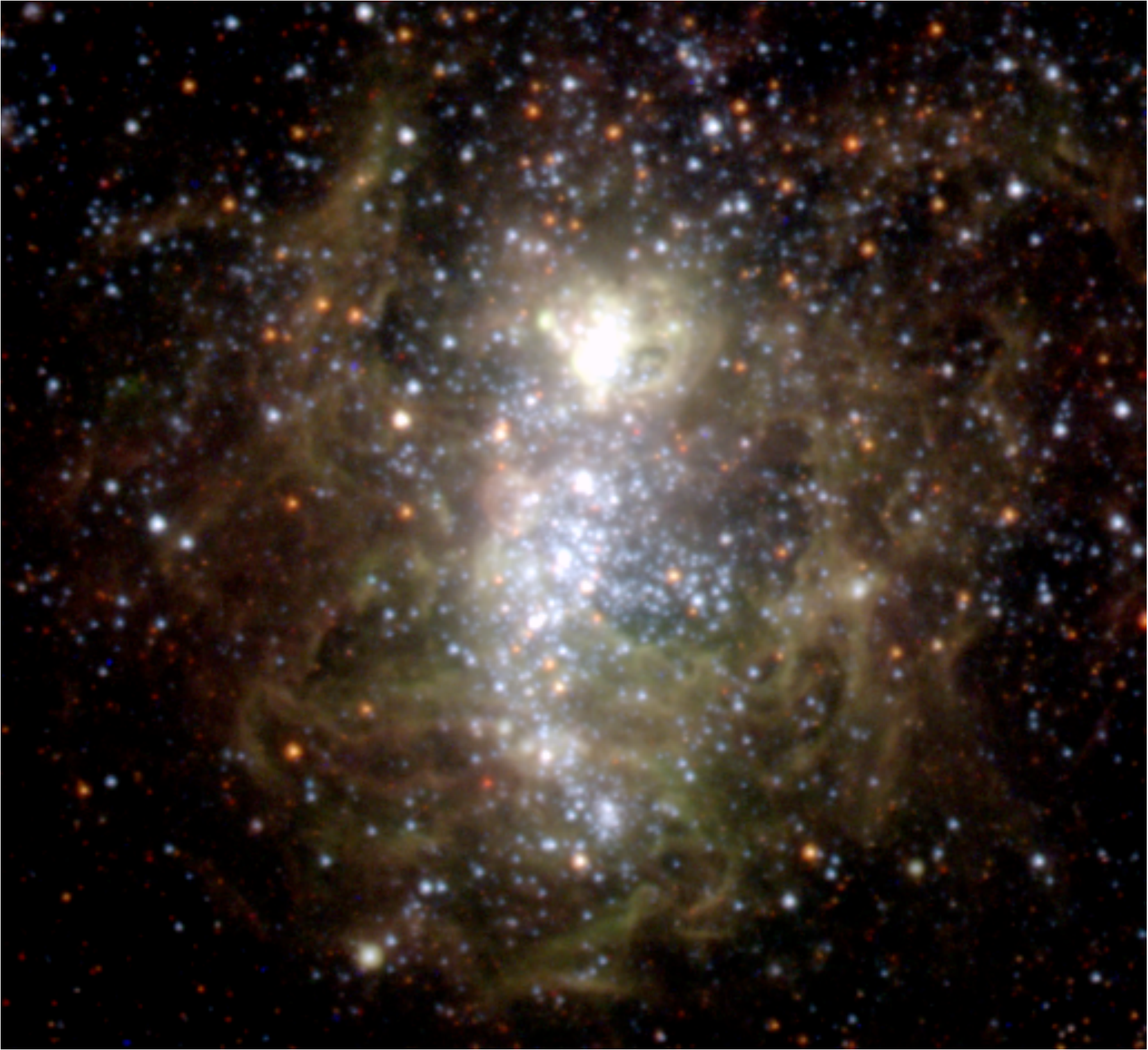}
		\caption{Colour composite of NGC~5455, obtained from {\em HST} WFC3 images in the F438W (blue channel), F555W (green channel) and F600LP (red channel ) filters. The field of view is equivalent to the KCWI data cube coverage, corresponding to $690\times630$~pc$^2$.}\label{HST5455}
	\end{minipage}\qquad
	\begin{minipage}[t]{.45\textwidth}
		\includegraphics[width=\columnwidth]{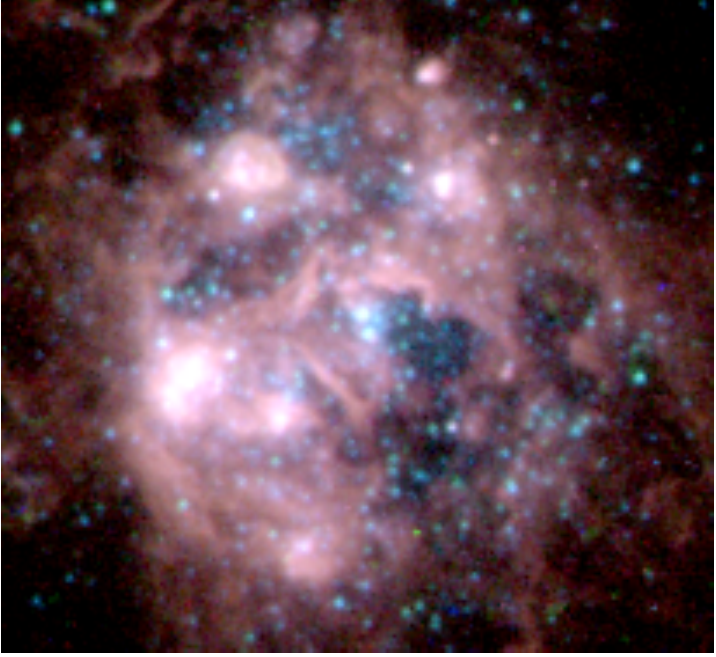}
		\caption{Colour composite of NGC~5471, obtained from {\em HST} WFPC2 images in the F547M (blue channel), F675W (green channel) and F656N (red channel ) filters. The field of view is equivalent to the KCWI data cube coverage, corresponding to $690\times630$~pc$^2$.}\label{HST5471}
	\end{minipage}
\end{figure*}

In Fig.~\ref{HST5455} and \ref{HST5471} we present high spatial resolution colour composite images of NGC~5455 and NGC~5471, respectively, that we obtained from {\em Hubble Space Telescope} ({\em HST\,}) archival data (Programs IDs 14678 and 6829 -- for colour images of NGC~5471 see also \citealt{Chen:2002, Chen:2005a}). The detailed morphological information they provide is helpful to interpret the kinematical data presented in the following sections. 
The color-magnitude analysis based on the NGC~5471 {\em HST} data by \citet{Garcia-Benito:2011} 
indicates that the star formation activity has proceeded at a nearly constant rate during the past $\sim$100~Myr over the extent of this giant \hii\ region.

\subsection{Global line profiles}\label{Sec:globalprofiles}
In order to compare the widths of integrated line profiles obtained from our data with published values we simulated 
observations through a $2'' \times 2''$ aperture, centred on knot A in both objects (see Fig.~\ref{n5455nb}\,--\,\ref{n5471nb}), and measured the FWHM of the \hbeta\ and \oiii\lin5007 lines from the extracted spectra, using Gaussian fits. We repeated the measurements with a 
$2'' \times 4''$ aperture, centred on the approximate geometrical centre of the ionized gas distribution, and finally using the whole field of view. 
In the latter case the velocity dispersions, referring to a $\sim$0.7\,kpc physical scale, can be directly compared to those obtained for poorly resolved systems (\eg\ \hii\ galaxies) viewed at large distances.

The observed velocity dispersion $\sigmaobs$, calculated as FWHM/2.355, was corrected for the effects of instrumental ($\sigmainstr$), thermal ($\sigmath$) and fine structure ($\sigmafs$ -- \hbeta\ line only) broadening, according to:

\begin{equation}
\sigma =  (\sigmaobs^2 - \sigmainstr^2 - \sigmath^2 - \sigmafs^2)^{1/2}.
\end{equation}

\noindent
We adopted $\sigmainstr = 7.6$~\kms\ (see Sec.~\ref{sec:observations}) and $\sigmafs$\,(\hbeta) = $2.4$\,\kms\ from \citet{Garcia-Diaz:2008}. The thermal component was evaluated, assuming a Maxwellian velocity distribution, as:

\begin{equation}\label{Eq:thermal}
\sigmath =  \left( \frac{k\,T_e}{m} \right)^{1/2},
\end{equation}

\noindent
where $m$ is the mass of the ion (either H$^+$ or O$^{++}$). For the electron temperature we referred to the $T$(\oiii) values published by \citet{Kennicutt:2003}, adopting \te\,=\,10,000~K for NGC~5455 (=\,H409 in their Table~2) and the average of four knots (A\,--\,D), \te\,=\,13,275~K, as representative for NGC~5471.

The results of this procedure are summarized in columns 3 ($2'' \times 2''$ aperture), 4 (geometrical center) and 5 
(full field of view) of Table~\ref{tab:comparison}. Our comparison data 
are taken from \citet[=\,SW70, column 6]{Smith:1970}, \citet[=\,H86, column 7]{Hippelein:1986} and \citet[=\,F18, column 8]{Fernandez-Arenas:2018}. 
We measure larger values of the velocity dispersion in NGC~5471 compared to NGC~5455, in agreement with H86, but in contrast with SW70 and F18.
Our $\sigma$ values for NGC~5455 are in good agreement with H86 if we consider the $2'' \times 2''$ aperture, but in the case of NGC~5471 we find agreement by selecting the other apertures.
The line widths reported in Table~\ref{tab:comparison} can be rather sensitive to the position chosen for the aperture center, both 
for our data and the published data, and we attribute the differences with published measurements to variations in aperture sizes and positions. As already observed by F18, such differences can amount to approximately 10\%, consistent with our findings.

\begin{table*}
	\centering
	\caption{Line width comparison.}\label{tab:comparison}
	\begin{tabular}{cccccccc}
\hline
			& 			&	\multicolumn{6}{c}{$\sigma$ (\kms)} \\
\hii\ region& line		&	$2'' \times 2''$ &	$2'' \times 4''$	&	full field			& SW70	&	H86		&	F18\\
			& 			&	aperture		 &	aperture			&	of view 			&    		&				&		 \\		
(1)			& (2)		&	(3)		 		&	(4)					&	(5)		 			& (6)   	&	(7)			&	(8)      	 \\								       
\hline
NGC~5455	& \hbeta	&	$20.2 \pm 0.4$$^1$	&	$21.8 \pm 0.4$$^3$	&	$21.2 \pm 0.3$	& 23.9		&	$20.6 \pm 0.4$	&	$23.4 \pm 1.1$  	\\
			& \oiii		&	$17.8 \pm 0.4$$^1$	&	$21.4 \pm 0.4$$^3$	&	$19.5 \pm 0.3$	& ...		&	$17.5 \pm 0.3$  &	...      	 		\\[2mm]		
NGC~5471	& \hbeta	&	$24.2 \pm 0.9$$^2$	&	$19.6 \pm 0.9$$^3$	&	$22.7 \pm 0.3$	& 19.4		&	$22.4 \pm 0.4$	&	$20.0 \pm 0.9$ 		\\
			& \oiii		&	$23.3 \pm 0.9$$^2$	&	$18.6 \pm 0.9$$^3$	&	$22.0 \pm 0.3$	& ...		&	$19.5 \pm 0.3$	&	...    				\\[-0.5mm]		
\hline
	\end{tabular}
	\begin{minipage}{13cm}
	$^1$Aperture centred on NGC~5455\,A.\,\,\,	$^2$Aperture centred on NGC~5471\,A.\,\,\, $^3$Aperture centred 
	on approximate geometrical centre of ionized gas distribution.\\
	SW70: \citet[taking $T_e = 10^4\,K$]{Smith:1970}; H86: \citet{Hippelein:1986}; F18: \citet{Fernandez-Arenas:2018}.
	
\end{minipage}
\end{table*}

\section{Single line profile fits}
\begin{figure*}
	\center
	\includegraphics[width=0.9\textwidth]{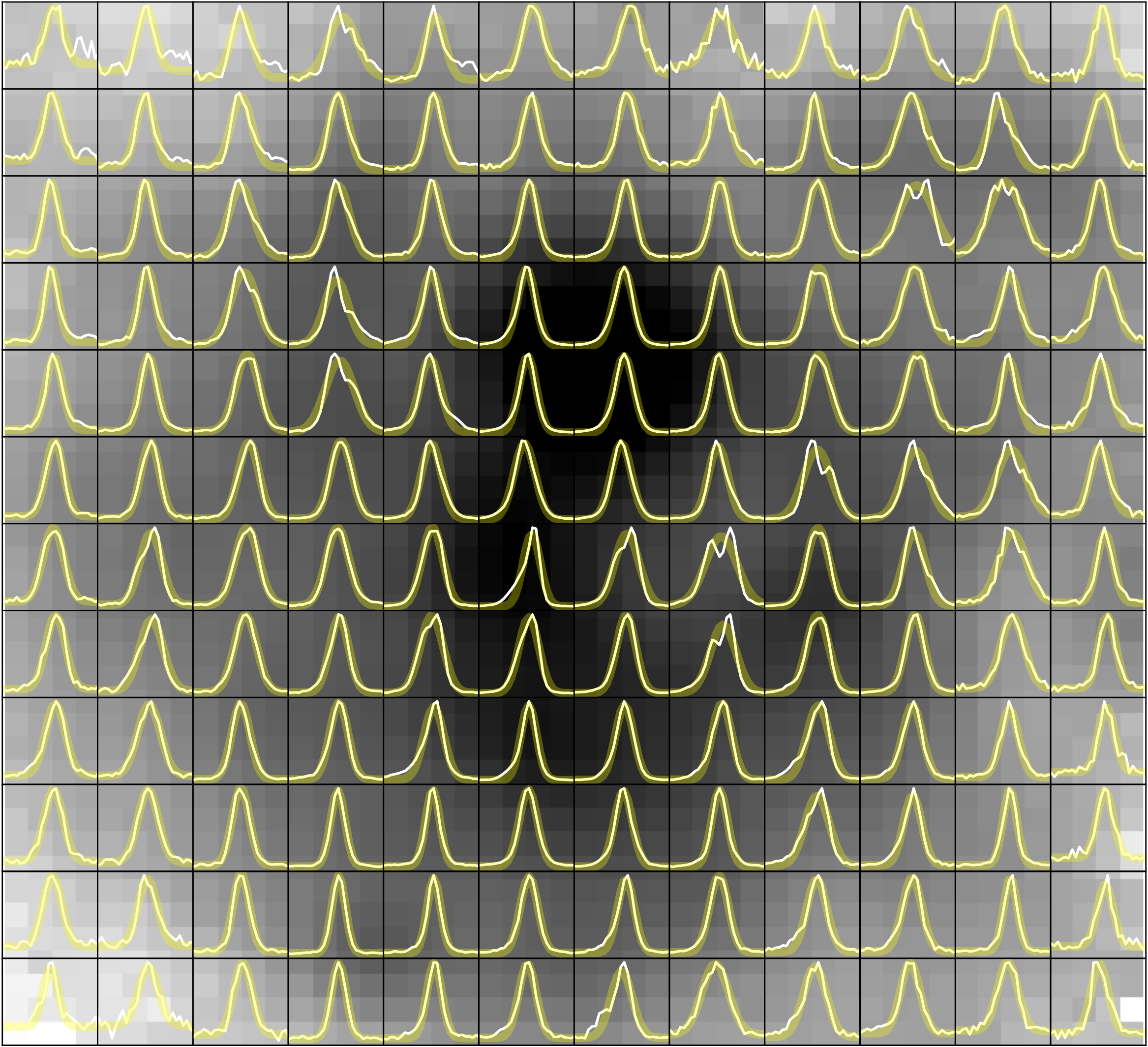}
	\caption{\oiii\lin5007 line profiles (white lines) and Gaussian fits (yellow thick lines) in individual spatial resolution elements across the \hbeta\ narrow-band image of NGC~5455. The spacing between elements is about 1.7 arcsec. In each subframe the radial velocity ranges between $-120$ and $+120$\,\kms, where the zero point is determined from the spatially integrated spectrum of the nebula.
	There is a large variation of peak intensity between the subframes, with strong lines approximating Gaussian curves, while fainter lines often display more complex profiles. 
		\label{n5455littlegaussians}}
\end{figure*}

\begin{figure*}
	\center
	\includegraphics[width=0.9\textwidth]{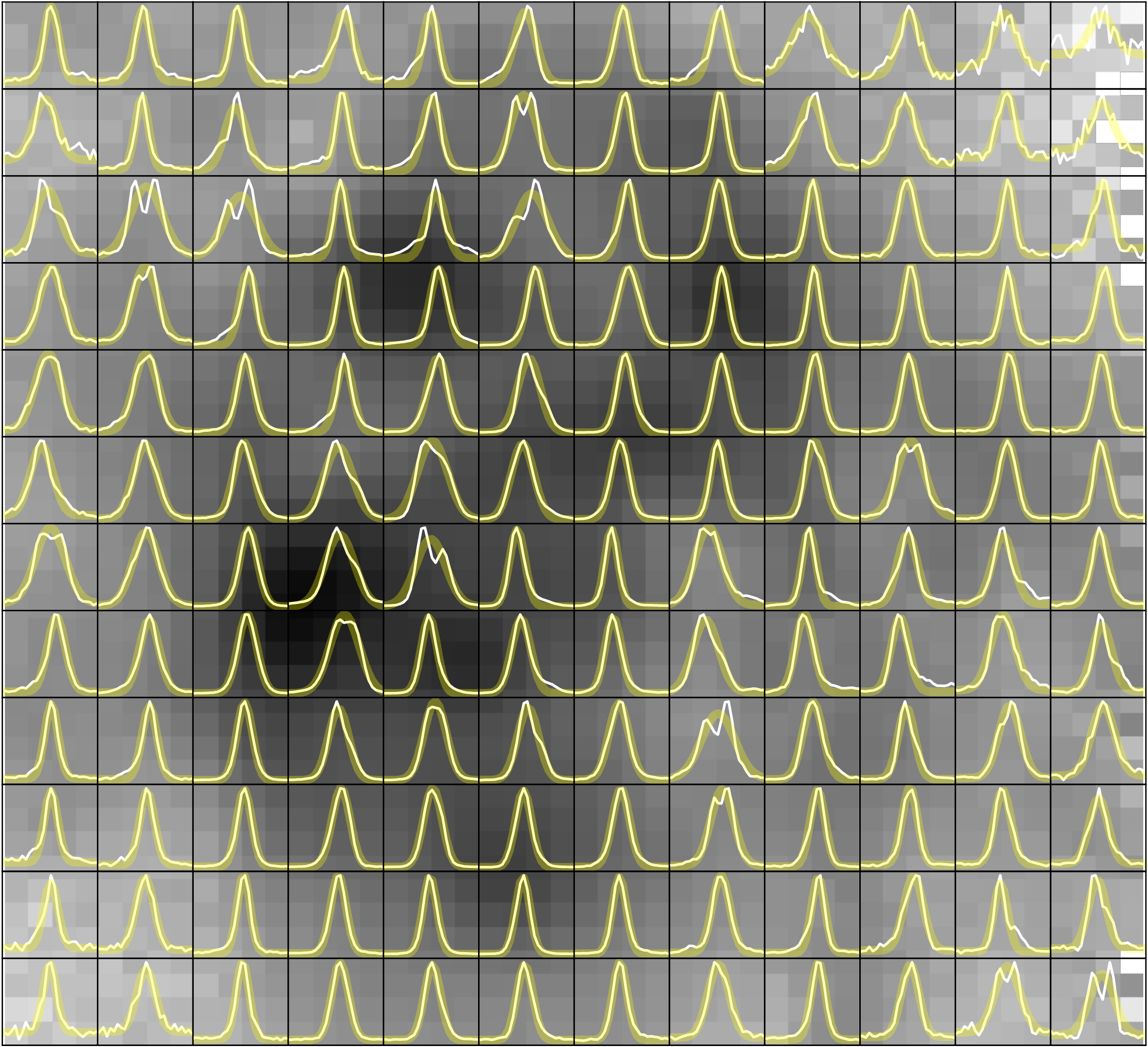}
	\caption{Same as Fig.~\ref{n5455littlegaussians}, for NGC~5471.
		\label{n5471littlegaussians}}
\end{figure*}

In this section we look at the spatially resolved behaviour of the line profiles across NGC~5455 and NGC~5471, by fitting single Gaussians to the profiles of both \hbeta\ and \oiii\lin5007\ in each spaxel. The line profiles of giant \hii\ regions often deviate from simple Gaussian curves, displaying asymmetries, extended wings and multiple components, indicative of 
the complex kinematics resulting from the superposition of discrete broken filaments and expanding shells along the line of sight (\citealt{Sabalisck:1995, Yang:1996}).
Numerous examples can be seen in Fig.~\ref{n5455littlegaussians} and \ref{n5471littlegaussians}, where we illustrate the variation of the \oiii\lin5007 line profile across the KCWI field of view in NGC~5455 and NGC~5471, respectively (we selected the stronger \oiii\lin5007 line in order to maximize the signal-to-noise ratio).
However, a single Gaussian fitting approach keeps the line profile analysis  easily tractable, at the same time providing important first-order information about the gas kinematics, via the determination of line width and velocity variations across nebulae. 
Fits to line profiles that are clearly split into at least two components are affected by large residuals, but still give insight into the presence of separate shells or filaments and their velocities  (\citealt{Munoz-Tunon:1996}).
A multi-component analysis that further addresses this issue will be presented in the following section.

The line profiles in each resolution element of the rebinned data cubes were fitted with single Gaussian functions with the \mbox{\tt lmfit} package (v. 0.9.13, \citealt{Newville:2019}).
For our analysis we retained lines whose peak is at least 10 times larger than the rms 
noise of the continuum, as a tradeoff between accuracy of the fit parameters and spatial coverage completeness.
This criterion impacted low-flux regions along the edges of the field of view.
The flux-weighted average centroid velocity was adopted as the zero-point of the velocity scale. The yellow lines in Fig.~\ref{n5455littlegaussians} and \ref{n5471littlegaussians} show examples of the Gaussian fits across both NGC~5455 and NGC~5471.

\subsection{$I-\sigma$ diagrams}\label{Sec:Isigma}
In two separate studies of the spatially resolved kinematics of the giant \hii\ regions NGC~604 and NGC~588 in the nearby galaxy M33, 
\citet[\,=\,M96]{Munoz-Tunon:1996} and \citet[\,=\,Y96]{Yang:1996} correlated the peak intensity ($I$) of single Gaussian fits to strong nebular lines
with either the velocity dispersion ($\sigma$) or the FWHM of the line profiles, respectively. The $I-\sigma$ diagram thus introduced represents a diagnostic tool that helps to differentiate emission line broadening mechanisms in giant \hii\ regions, and its use has been extended to kinematical investigations of \hii\ galaxies and other star-forming dwarf galaxies (\citealt{Martinez-Delgado:2007, Bordalo:2009, Moiseev:2012}). 
Based on the line widths they measured, both M96 and Y96 concluded that the gas motions are supersonic\footnote{The isothermal sound speed for an \hii\ region at $10^4$\,K is $c_s \simeq11$\,\kms\ (\eg\ \citealt{Spitzer:1968})} at most locations, 
but that the scatter in $\sigma$ increases strongly towards lower line intensities. The highest $\sigma$ values are encountered in the faintest nebular regions. Both papers associated the fainter and high-velocity dispersion regions with gas shell structures originating from the effects of stellar winds and SNRs, while concluding that virial motions are responsible for slower, but still supersonic, gas motions.
A minimum ($\sigma\simeq 17$\,\kms) value of the velocity dispersion was found for NGC~604, irrespective of line intensity, while in the case of NGC~588 a similar minimum could not be established, with a significant fraction of the nebular gas exhibiting subsonic  $\sigma$ values. This was interpreted as possibly due to a more evolved kinematic state. M96 demonstrated how prominent shell structures define inclined bands in the $I-\sigma$ diagram, extending from low $I$ and high $\sigma$ locations (corresponding to the shell centers -- the line width of the Gaussian fits measuring the maximum expansion velocity of the shells) to high $I$ and low $\sigma$ locations (at the shell edges).

\begin{figure*}
	\center
	\includegraphics[width=0.9\textwidth]{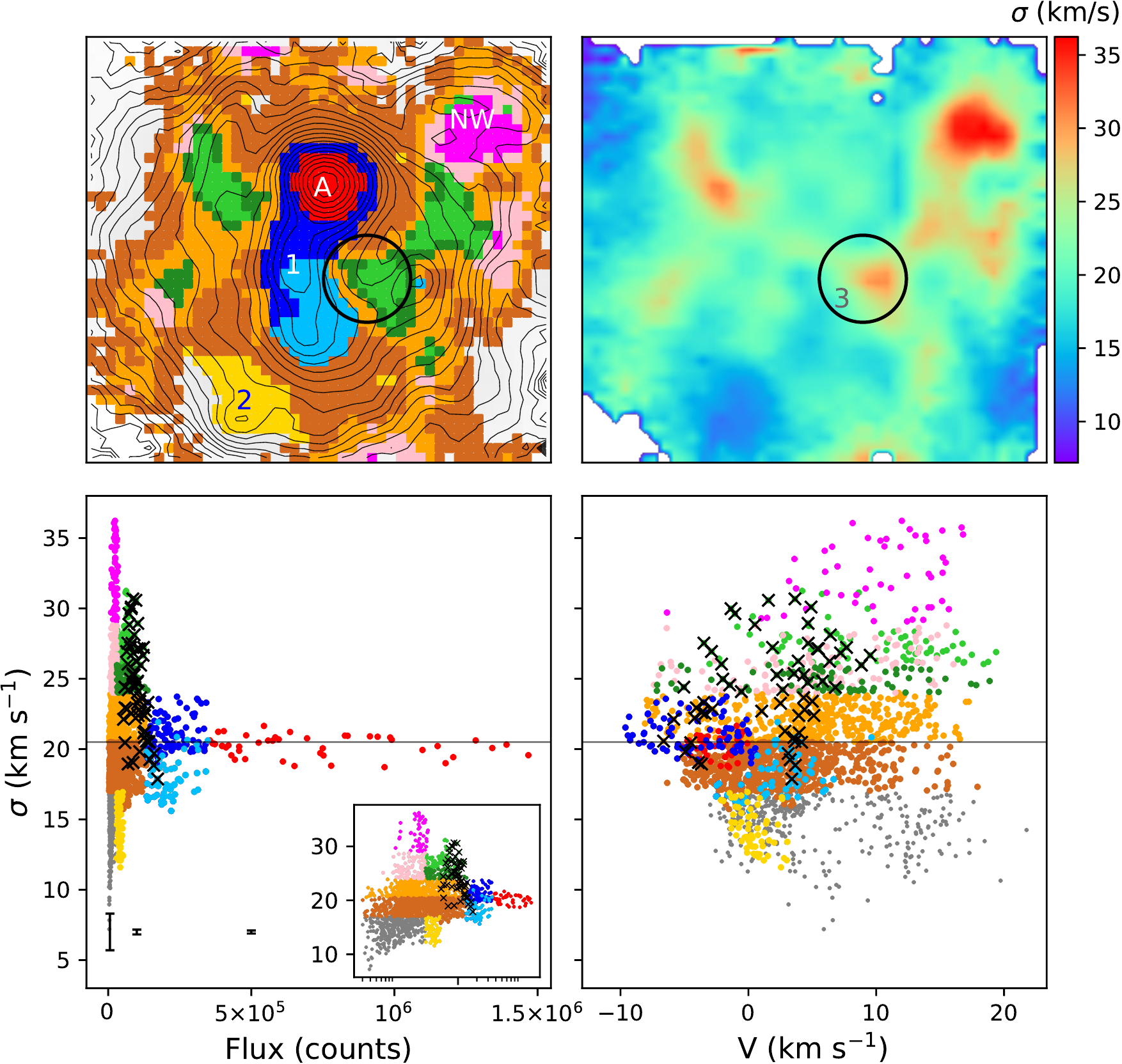}
	\caption{Results from single Gaussian fits for NGC~5455. (Top left): Isocontours of the \hbeta\ flux distribution (arbitrary level spacing). The colored areas represent regions selected in the $I-\sigma$ diagram shown in the panel below. (Top right): velocity dispersion map, smoothed with a bilinear interpolation. 
	(Bottom left): $I-\sigma$ diagram, where we use different colors to isolate portions that are mapped onto the top left panel. The error bars represent the uncertainties estimated by comparing the widths of the two \oiii\ lines, \lin4959 and \lin5007. The inset uses the logarithmic scale for the flux.
	(Bottom right): $V-\sigma$ diagram, using the same colour code. The horizontal line in the two bottom plots represents the flux-weighted average $\sigma$ value.
	The circle in the top row identifies the location of slit profiles with a maximum in $\sigma$, that we associate with the presence of a shell.
		\label{isigma5455}}
\end{figure*}

\begin{figure*}
	\center
	\includegraphics[width=0.9\textwidth]{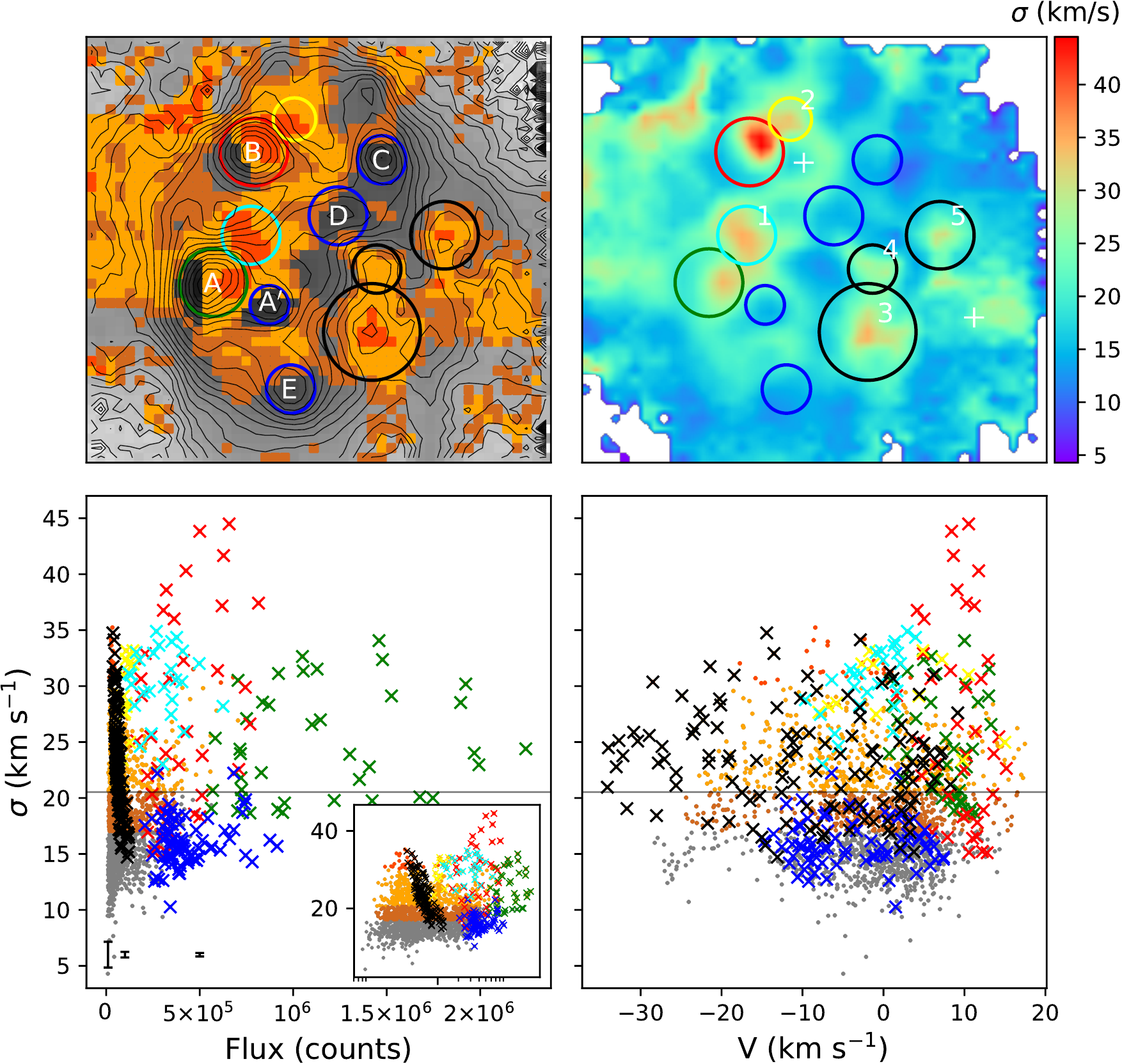}
	\caption{Same as Fig.~\ref{isigma5455}, for NGC~5471. See text for the identification of the different features.
		\label{isigma5471}}
\end{figure*}

In the bottom-left panels of Fig.~\ref{isigma5455} and \ref{isigma5471} we display the $I-\sigma$ diagrams for NGC~5455 and NGC~5471, respectively, obtained from fits to the \hbeta\ line (equivalent information is gathered from the \oiii\lin 5007 line). Instead of the peak intensity we have used the fitted line
flux -- this choice introduces no significant difference.
In each figure we use colors in order to map different regions of the 
$I-\sigma$ diagram to the corresponding emitting location in the top left panel (where \hbeta\ image isophotes are displayed). The error bars at the bottom represent the standard deviation of the difference $\sigma$(\oiii\lin5007) $-$ $\sigma$(\oiii\lin4959), an estimate of the uncertainty in $\sigma$, since the two lines have the same intrinsic width.
Maps of the velocity dispersion $\sigma$ are shown in the top right panels. Finally, the $V-\sigma$ diagrams in the bottom right panels display the velocity centroid of each Gaussian as a function of the velocity dispersion. The latter diagram tests whether portions of the nebulae possess streaming rather than random motions along the line of sight 
(see \citealt{Martinez-Delgado:2007, Bordalo:2009}).
The two bottom diagrams include a horizontal line, drawn at the flux-weighted average $\sigma$ value: 20.6 and 20.7\,\kms\ for NGC~5455 and NGC~5471, respectively. These differ somewhat from those reported in column 5 of Table~\ref{tab:comparison}. This is due to the fact that in that case we used the full (unbinned) data cube information. 

\subsubsection*{NGC 5455}
The $I-\sigma$ diagram of NGC~5455 (Fig.~\ref{isigma5455}, bottom left panel) appears qualitatively similar to the diagram presented for NGC~588 by M96, where an approximately constant lower $\sigma$ is {\em not} observed.

We have assigned different colors to various regions in this diagram, in order to identify (with corresponding colors in the top left panel) where the emission originates. The main emission knot (labelled `A' -- red color) defines the bright tip of the main horizontal band in the $I-\sigma$ diagram, with  $\sigma\simeq 18-21$\,\kms. The largest $\sigma$ values in NGC~5455 are encountered in a slightly offset position from a weak  filament labelled 'NW', and are represented in magenta. In contrast, a weak emission peak, labelled `2', corresponds to a region of low $\sigma$ (in yellow), well confined in the $I-\sigma$ plot. The bright plume to the south of the main knot, labelled `1' (blue and cyan), displays a small range in velocity dispersion. The $V-\sigma$ diagram in the bottom right panel indicates a mild velocity
gradient of $\sim$15\,\kms\ across the plume, with the northern (southern) portion having negative (positive) velocities.
Other regions of relatively large $\sigma$ ($24\,-\,30$\,\kms\ -- green and pink) appear spatially well defined and immersed in diffuse gas characterized by a dispersion similar to that of the  brightest blob 
($\sigma\sim 17-24$\,\kms\ -- orange and maroon). 
Knot A appears to be encircled by a high-$\sigma$ network (in green). 
Towards the edges of the field of view the velocity dispersion falls below 15~\kms\ (in grey in the left panels), and some of the faintest points approach or fall below the isothermal sound speed. 

The {\em HST\,} image of Fig.~\ref{HST5455} shows a cavity ($\sim$100\,pc in diameter) in the ionized gas distribution, immediately to the W of the bright plume, and centred on an ensemble of well-resolved bright stars. At this location 
we observe split line profiles (see Fig.~\ref{n5455littlegaussians}) and
find a peak in the $\sigma$ map, identified by a circle labelled as `3' in Fig.~\ref{isigma5455} (top right). The Gaussian fits within this circle define an inclined band in the  $I-\sigma$ diagram (black crosses in the bottom left panel). 
The association of a visible shell structure with such an inclined band agrees with the findings of M96. If we interpret the  maximum measured velocity dispersion ($\sim$30\,\kms) as the velocity separation between receding and approaching sides of the shell along the line of sight, we obtain a dynamical age for this  structure of 3.3~Myr. 	
	
Several arcs and filaments are present throughout the region covered by our KCWI observations. In particular what appears to be a faint, large broken shell, with a diameter of $\sim$160\,pc, is barely visible in the NW corner of the {\em HST\,} image, matching the location where we measure the largest velocity dispersion ($\sim$35\,\kms) in the NGC~5455 area. We derive a dynamical age of approximately 4.5~Myr, but, contrary to the smaller shell we have identified earlier, there is puzzlingly no prominent association of bright stars in this region: we cannot explain this localized large gas turbulence with the presence of massive stars.
It is conceivable that we are observing a blowout region (\citealt{Mac-Low:1988, Tenorio-Tagle:1988}) venting gas into the galactic halo.

\subsubsection*{NGC 5471}
The case of NGC~5471, a multi-core giant \hii\ region (\citealt{Kennicutt:1984}), appears more complex compared to NGC~5455. We can recognize several different features in the narrow band images, identified by circles in the top left panel of Fig.~\ref{isigma5471} (including the knots we labelled in Fig.~\ref{n5471nb}). Points inside these circles are mapped onto the $I-\sigma$ diagram (bottom left panel) with crosses of the same color. The single, most striking feature of the diagram is that large values of the velocity dispersion are met throughout the intensity scale. In particular, the largest $\sigma$ is measured in the luminous knot B (red circle and crosses), where the $\sigma$ range is also very high (15\,--\,45\,\kms). These findings are largely due to the inadequacy of a single Gaussian fit for this object, as we will describe in Sect.~\ref{Sec:knotB}. In fact, the line profile in this area (see Fig.~\ref{n5471littlegaussians}) shows a faint, but very broad, component superposed on a stronger, narrow component, matching the profile shape presented by CK86.
The $\sigma$ map in the top right panel shows that it is the NW sector of the area encircled for knot B that is characterized by a high velocity dispersion, up to $\sim$45\,\kms. This roughly agrees with the location where a large \sii/\halpha\ line ratio is found in the {\em HST} imaging presented by \citet{Chen:2002}, and associated with the SNR shock region. Further details will be presented in Sect.~\ref{Sec:knotB}.

The velocity dispersion is high (20\,--\,35\,\kms) also in knots A (green), and in two small regions to the NW of both knot A (`1' -- cyan) ad knot B (`2' -- yellow), labelled in the top right panel for clarity. 
We note that also in the case of knot A, as in knot B, the highest $\sigma$ values are confined to a small area to the W of the knot center.
Interestingly, the other knots (A$^\prime$, C, D and E), are characterized by {\em low} velocity dispersion values
$\sigma < \sigma_0$, where $\sigma_0 = 20.7$\,\kms\ is the flux-weighted average velocity dispersion.
We have represented them using the same blue color, because of the similar location they occupy in the $I-\sigma$ diagram, although the $V-\sigma$ plot (bottom right) shows a range of radial velocities.

We display in orange (red) the remaining points with $\sigma_0 < \sigma < 30$ ($>30$) \kms. The low-$\sigma$ knots (from E in the south to C in the north) are located in a low-$\sigma$ ring-like structure surrounding knots A and B, that roughly matches the edges of the elongated ovoidal bubble easily recognizable in narrow-band images (Fig.~\ref{n5471nb}). We speculate that 
we might be observing a rough $\sigma$-cluster age relation in this giant \hii\ region. The {\em HST} data analysis by \citet{Garcia-Benito:2011} led to an estimated age of $3\pm2$ Myr for the ionizing cluster in knot A, and $5\pm2$ for the clusters in knots A$^\prime$, B, C, D and E, although the authors caution about the results for the fainter objects (C, E). Moreover, for two clusters whose position is indicated by crosses in the top right panel \citet{Garcia-Benito:2011} derive an even older age of $9\pm3$~Myr, while the velocity dispersion in this area is around 20\,\kms.
There is thus some circumstantial evidence that a young cluster (A) is associated with higher-than-average velocity dispersion, while for older clusters $\sigma$ falls back to average or below-average values.

To the west lies a region where the velocity dispersion rises again, and where we identified three possible expanding shells, indicated by black circles and labelled with numbers 3, 4 and 5. The points in these areas define an inclined band in the $I-\sigma$ diagram, the signature of shell structures in the interpretation given by \citet{Munoz-Tunon:1996}.  The spaxels contained within the three putative shells span a wide range in velocity centroid ($-30$ to 10\,\kms\ -- see bottom right panel). These structures correspond to roughly circular `clearings' in the ionized gas distribution in the {\em HST\,} image of Fig.~\ref{HST5471}. The largest one (`3', $\sim$70\,pc in radius), following the same interpretation of an expanding shell as done before, yields a kinematic age of $\sim$4~Myr.

\subsection{The radial velocity field}
\begin{figure}
	\center
	\includegraphics[width=\columnwidth]{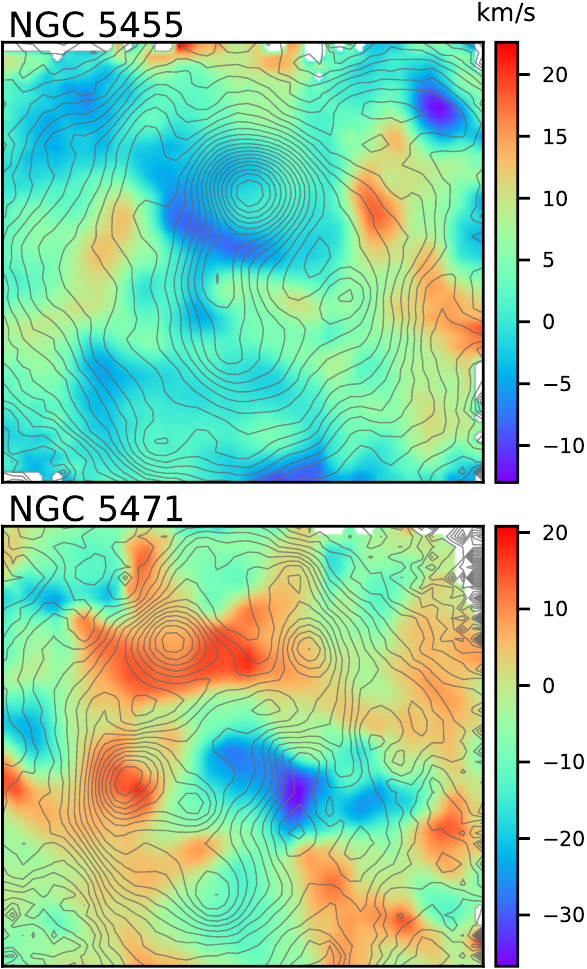}
	\caption{Radial velocity field in NGC~5455 (top) and NGC~5471 (bottom) from single Gaussian fits to the \oiii\lin5007 line. In this and subsequent plots we include arbitrarily spaced isophotes of the \hbeta\ emission, for reference.
		\label{velocitymap}}
\end{figure}

In Fig.~\ref{velocitymap} we display maps of the velocity centroids for our two targets, determined from single Gaussian fits to the \oiii\lin5007 line (we remind the reader that the velocity zero point is defined by the flux-weighted average velocity). The radial velocity range in NGC~5471 is $\sim$60\% higher than in NGC~5455, indicative of the larger gas motions in the higher luminosity, multi-core nebula.
We ascribe the morphology of these maps to the filamentary distribution of the ionized gas and the presence of expanding shells. Different filaments or 
`voids' that we can identify in the {\em HST\,} images can be matched with features in the velocity field. For example, the significant radial velocity gradient ($\Delta V \simeq 30$\,\kms) we observe in the NW region of NGC~5455 develops across the broken bubble we described earlier, with filaments at the top approaching us at 10-12\,\kms, those near the bottom receding at 15-18\,\kms. This is also the location with the highest $\sigma$ in Fig.~\ref{isigma5455}.
The arc of negative velocity immediately to the SE of knot A corresponds to a transition band of lower stellar density  identifiable in the 
{\em HST\,} image (Fig.~\ref{HST5455}) between knot A and the bright stellar clusters to its south.
For NGC~5471 we can associate the largest approaching radial velocity with the intersecting region between shells 3 and 4 (se Fig.~\ref{isigma5471} for identification). The gas in the `shell area' is generally characterized by negative velocities (\ie\ approaching the observer), while the gas surrounding knots A and B moves at the highest receding velocity.

\section{Multiple line profile fits}\label{Sec:MultiGauss}
The emission line profiles of nearby giant \hii\ regions and blue compact galaxies often deviate from single Gaussians, and multiple narrow subcomponents are revealed at high spectral resolution (\citealt{Chu:1994, Melnick:1999, Bordalo:2011, Chavez:2014}). 
In addition, a relatively low spectral resolution is sufficient to uncover a broad underlying component  
(\citealt{Mendez:1997, Firpo:2011, Hagele:2013}), from which expansion velocities of several hundred \kms\ are inferred, although extreme values of thousands of \kms\ have also been measured (\eg\ \citealt{Castaneda:1990, Roy:1992}). A broad underlying component is also observed in non-giant \hii\ regions (see \citealt{Castaneda:1988} for the case of the Orion nebula).

Multiple Gaussian fits were calculated for each spaxel of our KCWI data using 
\mbox{\tt lmfit} (\citealt{Newville:2019}), allowing a number of components up to $\ncomp = 8$, 
if statistically justified. This decision was based on the Bayesian information criterion (BIC -- \citealt{Schwarz:1978}), which is widely used in astrophysics for model selection (\citealt[]{Liddle:2004}). We preferred the BIC over the Akaike information criterion (\citealt{Akaike:1974}), also calculated by \mbox{\tt lmfit}, because it is more conservative in the number of model parameters (in our case the number of Gaussians to include).
We followed guidelines from \citet[see also \citealt{Liddle:2007} or \citealt{Shi:2012} for astronomy-related examples]{Kass:1995} in adopting a minimum decrease $\Delta BIC = -10$ to justify the inclusion of an additional Gaussian component in our model profiles.

In the case of \oiii\lin5007, the strongest line in our data, the majority of our fits (about 3/4 of the spaxels) are 
characterized by $2 \leq \ncomp \leq 4$.
We experimented with forcing the weaker \hbeta\ lines to have the same number of components at the same wavelengths as the \oiii\lin5007 lines, allowing for different widths and intensities. We found that this procedure works in some cases, but more often the comparison of the decomposition of the two line profiles is less than satisfactory -- the significantly larger thermal broadening of the Balmer lines is the culprit. The different amounts of thermal broadening explains why at the very high spectral resolution ($R \sim 10^5$) used by \citet{Melnick:2019} the \oiii\ lines are resolved into a larger number of components than \hbeta.
Besides, trying to match the components in the two lines worsens the fit to the \hbeta\ line compared to an unconstrained fit, and consequently we abandoned this strategy.

In the following analysis we adopt a conservative approach, in order to avoid the risk of over-interpreting the fits: the line decomposition is not unique, and depends on the signal-to-noise ratio and the spectral resolution of the observations.
Our principal aims are to ascertain the presence of an underlying broad component, measure its extent, and quantify the velocity scatter between the different gas clouds that we are able to identify along the line of sight.


\begin{figure}
	\center
	\includegraphics[width=\columnwidth]{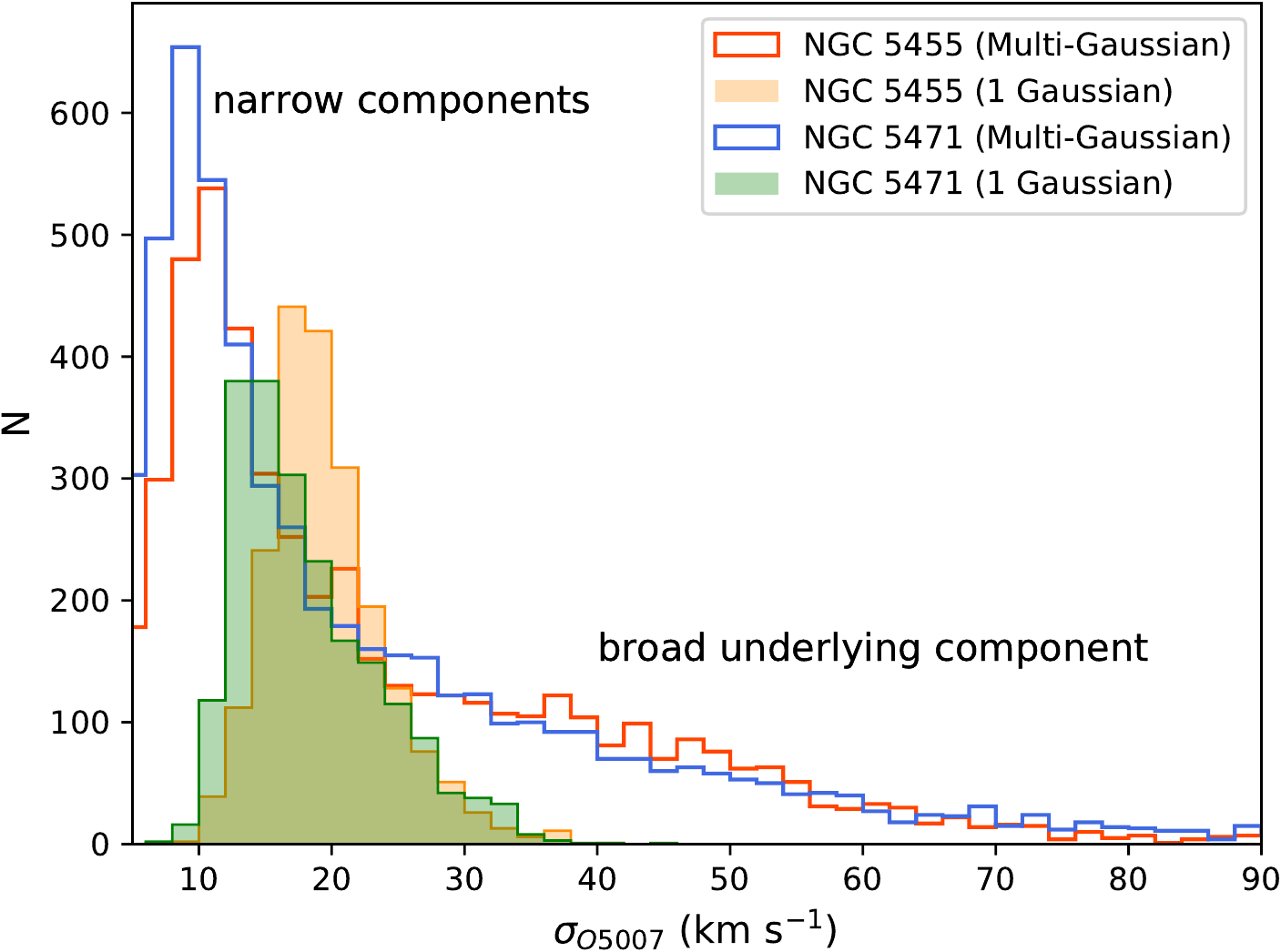}
	\caption{Histograms of the \oiii\lin5007 velocity dispersion from the multiple Gaussian analysis of NGC~5455 (red curve) and NGC~5471 (blue curve). The narrower histograms from the single Gaussian fits are also shown: NGC~5455 (solid yellow) and NGC~5471 (solid green).
		\label{histogram}}
\end{figure}

In Fig.~\ref{histogram} we display the distribution of the \oiii\lin5007 velocity dispersion (corrected for thermal and instrumental effects) for all components found in NGC~5455 (red curve) and NGC~5471 (blue curve). The two curves are quite similar, increasing steeply with decreasing $\sigma$, below 25\,\kms. The sharp drop seen at $\sigma < 8$\,\kms\ is likely due to our inability to measure line widths approaching our spectral resolution. The histograms obtained from the single-Gaussian fits (in solid  yellow and green) are much narrower. 
Values $\sigma > 25-30$\,\kms\ measured from the multiple Gaussian profile fits refer to underlying broad components found throughout the extent of the \hii\ regions, that cannot be uncovered by single Gaussian fits (see Sect.~\ref{Sec:broadcom}).
The high intensity portion of each profile is composed of narrower, often multiple, components, with $\sigma < 20-25$\,\kms. This dichotomy, that explains the presence of two different regimes in the multipple Gaussian histograms, is indicated in the plot.

Fig.~\ref{histogram} illustrates how a large fraction of the components resulting from our multiple Gaussian analysis exhibits line widths near or below the sound speed (around 11\,\kms), whereas 
single Gaussian fits are characterized by supersonic widths in virtually all cases.
At higher spectral ($\sigmainstr=3$\,\kms) and spatial (0.1~pc) resolution \citet{Melnick:1999} found that {\em all} the components they measured in 30 Dor have subsonic line widths, except for a pervasive, broad underlying one. They  concluded that 
the combination of multiple discrete filaments, shells and gas clumps, each characterized by a narrow velocity dispersion, is responsible, in the case of 30 Dor, for the supersonic line profile width observed at lower spectral resolution, in agreement with the conclusion drawn by \citet{Chu:1994}. In our study we observe supersonic narrow subcomponents, but we cannot discard the possibility that they, too, would break up into narrower, subsonic profiles, if observed at higher spectral and, importantly, spatial resolution, since our seeing-limited line of sight integrates the kinematic information over $\sim$30\,pc in the plane of the sky.
In fact, the line profile of each spaxel results from a combination of individual filament profiles and the velocity distribution of all the filaments whose light contributes to the spaxel. Spatial resolution plays therefore a pivotal role in disentangling the kinematic superposition of discrete gas elements. This affects the number of separate components revealed by the multiple profile analysis, and implies that averaging over large apertures (\eg\ for the derivation of global profiles -- see Sect.~\ref{Sec:global}) produces broad, supersonic profiles, even when the individual gas cloudlets are characterized by much narrower, subsonic line profiles. Ultimately, comparative conclusions on the widths of 
the line profiles of extragalactic H\,{\sc ii} regions need to take into account the vastly different physical resolutions attained by different studies: 1 arcsec corresponds to 33\,pc at the distance of M101, 4\,pc in M33 and only 0.25\,pc in the LMC.
In this context it is  worth mentioning that the work on NGC~604 by \citet{Yang:1996}, carried out at  higher spectral and physical spatial resolution than our present investigation, found supersonic line widths virtually everywhere in the nebula, based on {\em single} Gaussian fits. As illustrated above, only the nearby 30~Dor nebula has been entirely resolved into subsonic components.


\begin{figure*}
	\center
	\includegraphics[width=0.9\textwidth]{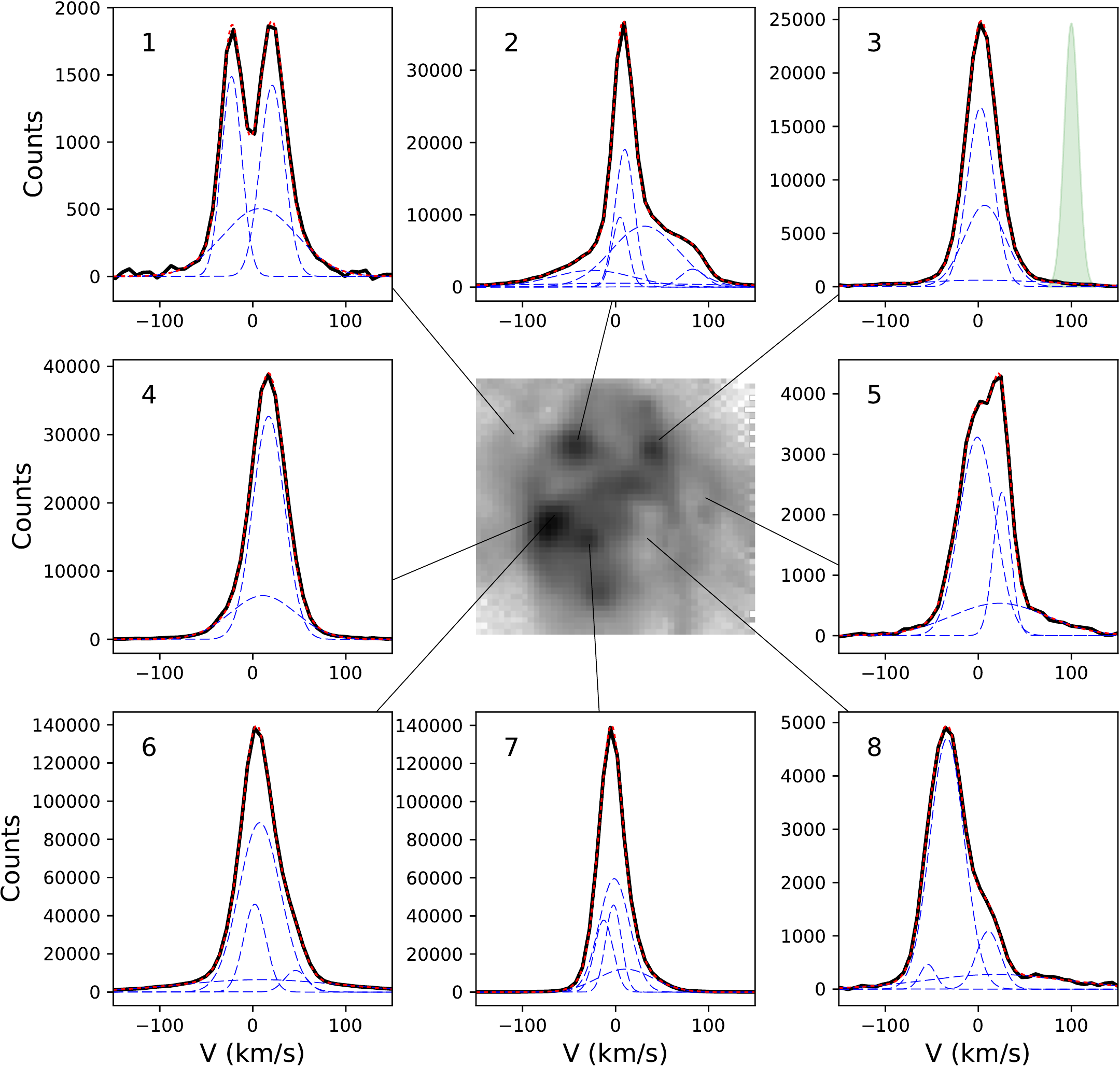}
	\caption{Representative examples of multiple profile fits to the \oiii\lin5007 line for NGC~5471. In each panel we display the observed line profile as a black continuous line, the model fit as a dotted red line, and the different model components as dashed curves. The green filled Gaussian at the right in panel 3 represents the instrumental profile.
		\label{profiles}}
\end{figure*}

We show representative examples of our multiple profile fits for NGC~5471 in Fig.~\ref{profiles}. We include clearly split or asymmetric line profiles (panels 1, 2, 5, 8), a spaxel in the SNR-hosting NGC~5471\,B (panel 2), line models that include a broad and faint component (panels 3, 6, 7, 8), and a simple profile modeled with only two components (panel 4). 
The asymmetry seen in panels 5 and 8, where an extended wing is detected at red wavelengths (positive velocities), is typical of the SW region of NGC~5471, which appears to be dominated by shell structures, as discussed in Sec.~\ref{Sec:Isigma}. In general, we find complex line profiles, decomposed into multiple Gaussians, in regions where we have evidence for the presence of shells. This is similar to what is observed, at much higher spatial resolution, in the nearby 30\,Dor (\citealt{Torres-Flores:2013}).


\subsection{Broad line components}\label{Sec:broadcom}
Broad line components have been observed in spatially resolved observations of extragalactic \hii\ regions and starburst galaxies (\citealt{Yang:1996, Melnick:1999, Westmoquette:2007, Bosch:2019}). 
These broad features (or wings) are likely due to stellar winds from massive stars and supernova explosions affecting the surrounding interstellar medium. 
\citet{Melnick:1999} speculated that the ubiquitous broad component ($\sigma\sim 45$\,\kms) they detected in the central parts of 30 Dor is the outcome of photo-erosion of dense gas clouds,
producing a diffuse and highly turbulent gas component. 
In a similar vein, \citet{Westmoquette:2007} proposed that the broad emission they measured for the starburst galaxy NGC~1569 ($\sigma$\,$\sim$\,40\,--\,120~\kms)
originates at the surface of cold clumps immersed in hot stellar winds. They invoked the presence of a turbulent mixing layer (\citealt{Begelman:1990}), formed at the interface between the dense clouds and the  stellar winds, which can give rise to emission lines featuring broad wings (\citealt{Binette:2009}). The detection of broad underlying line profiles ($\sigma$\,$\sim$\,20\,--\,65~\kms) at the edges of cold gas pillars in the vicinity of galactic ionizing clusters by \citet{Westmoquette:2010, Westmoquette:2013a} lends support to this interpretation.


A broad underlying component was detected for a large fraction ($\sim$2/3) of the line profiles we examined in NGC~5455 and NGC~5471, typically having 
$\sigma = 30-50$\,\kms, which is comparable to the widths found in 30~Dor (\citealt{Melnick:1999, Torres-Flores:2013}).
In certain regions of NGC~5471, namely in the vicinity of knots A, B and C, we measured considerably larger widths, $\sigma > 80-100$\,\kms. At these locations we detect very extended  
low-intensity profiles, examples of which can be seen in panels 2, 3 and 6 of Fig.~\ref{profiles}.


\subsubsection{Full width at zero intensity}
At each spaxel we calculated the Full Width at Zero Intensity (FWZI), defined by the velocity range within which the line profile lies above the rms scatter of the continuum (such a definition is subjective and dependent on the signal-to-noise ratio, making quantitative comparisons between different investigations difficult).
We compared the spatial distribution of the FWZI and the maximum $\sigma$ of the line profiles, which is dominated by the 
presence of the underlying broad component.
We found quite a good correlation, which indicates that even without performing a multi-profile fit one could use the FWZI as a proxy for the width of the broad component, thus  simplifying the profile analysis, and offering a method to investigate the presence of underlying broad components even utilizing lower spectral resolution and independent of the accuracy of a multi-component analysis.

\begin{figure}
	\center
	\includegraphics[width=0.9\columnwidth]{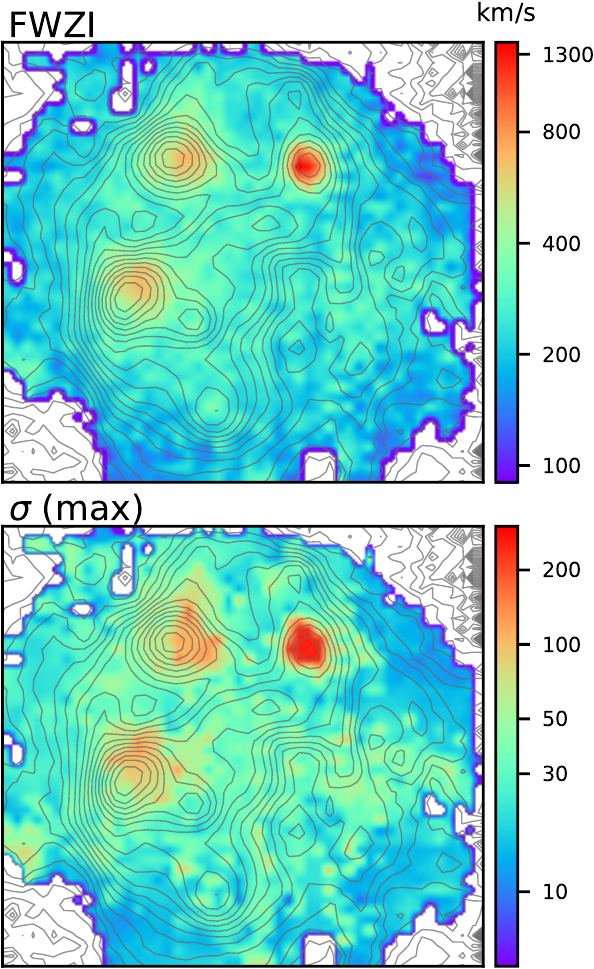}
	\caption{Spatial distribution of the FWZI (top) and maximum velocity dispersion (bottom) of the 
		\hbeta\ line for NGC~5471.
		\label{fwzi_vs_broad}}
\end{figure}

Fig.~\ref{fwzi_vs_broad} illustrates the case of the \hbeta\ line in NGC~5471. 
The FWZI can be somewhat affected by the presence of line splitting, which broadens the overall line profiles, but the effect is on the order of 50\,\kms, while we typically measure FWZI values of several hundred \kms, up to $>$\,1000\,\kms.
In our profile decompositions broad underlying components appear to have a minimum $\sigma \approx 25-30$\,\kms. Fig.~\ref{fwzi_vs_broad}  shows that the presence of the broad component occurs throughout the area we examined, except approaching the south and west edges of the field of view. Adopting the interpretation introduced at the beginning of this Section, the map therefore suggests that the influence of massive star winds on the turbulent motion of the gas in a giant \hii\ region is indeed ubiquitous, as found by \mbox{\citet[on a scale of 80~pc]{Melnick:1999}}, extending roughly 250\,pc from the centre of NGC~5471.

\begin{figure*}
	\center
	\includegraphics[width=0.65\textwidth]{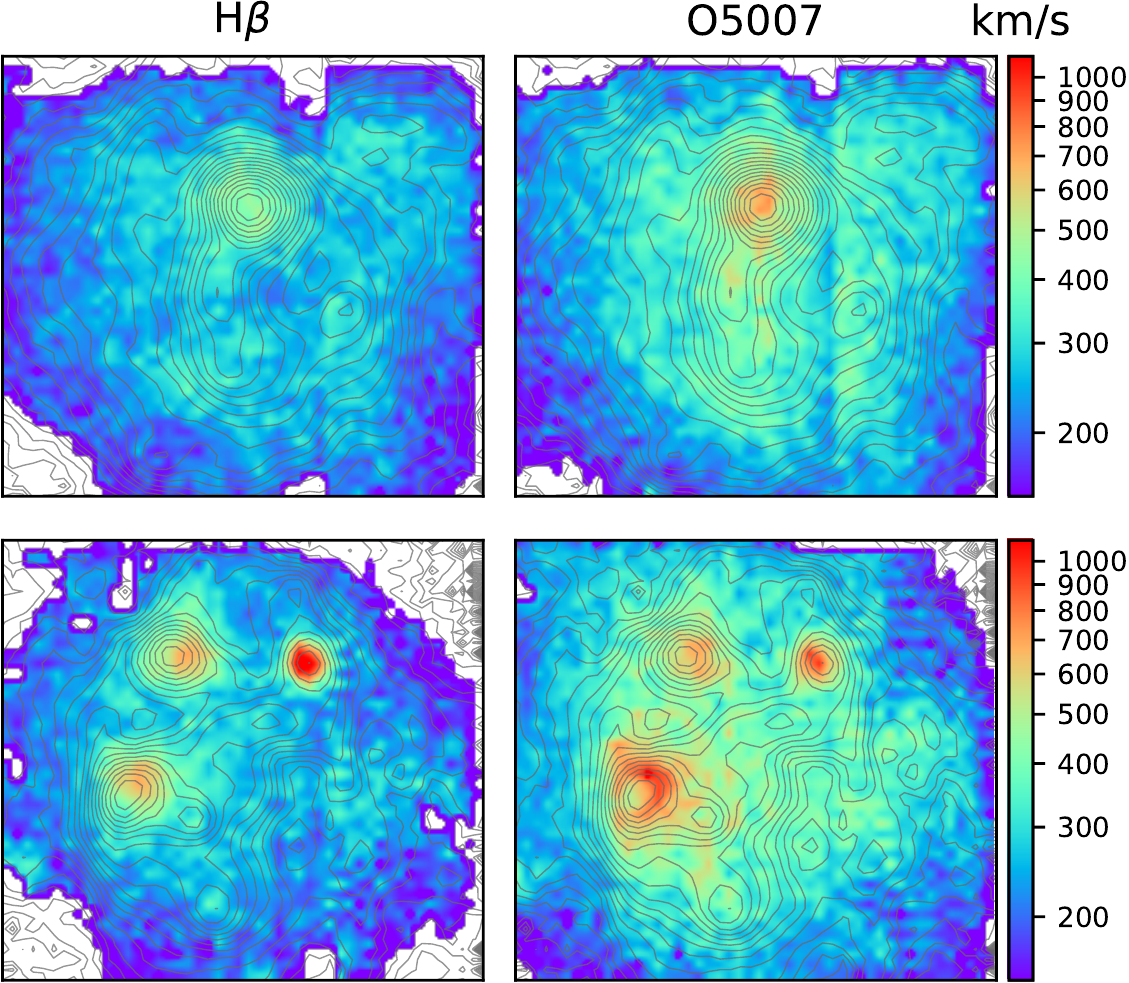}
	\caption{FWZI maps of the \hbeta\ and \oiii\lin5007 lines for NGC~5455 (top) and NGC~5471 (bottom).
		The same velocity range is used in order to facilitate the comparison between the two \hii\ regions.
		\label{FWZI}}
\end{figure*}


Maps of the FWZI for the \hbeta\ and \oiii\lin5007 lines in both our targets are displayed in Fig.~\ref{FWZI}. Their appearance is very different compared to the $\sigma$ maps presented in Fig.~\ref{isigma5455}\,-\,\ref{isigma5471}, obtained from single Gaussian fits. In regions located away from the main star-forming knots we measure FWZI $\simeq$ 300-500 \kms, while at the position of some of the main star-forming knots the FWZI can be much larger ($> 700$\,\kms). 


\subsubsection{Evidence for supernova remnants}
Observed at sufficiently high spectral resolution the line profiles of extragalactic SNRs appear broader than those of \hii\ regions (\citealt{Points:2019}). In most cases a narrow component (from unshocked \hii\ region gas -- $\sigma < 30$\,\kms) sits on top of a broad component (shocked gas) with $\sigma$ $\sim$ 40\,--\,100\,\kms\ (\citealt{Blair:1988, Dunne:2000}).
In the case of NGC~5471\,B, CK86 measured $\sigma$(broad) $\simeq$ 62\,\kms\ for \halpha. Our map in Fig.~\ref{fwzi_vs_broad}
indicates a much larger value for \hbeta, $\sigma$(broad) $\simeq250$\,\kms, because it refers to the much fainter extended component (visible -- for \oiii\lin5007 -- in panel 2 of Fig.~\ref{profiles}). The analysis of the much brighter broad component is presented in Sect.~\ref{Sec:knotB}.

In addition to NGC~5471\,B, CK86 found large low-intensity velocity widths in the line profiles of NGC~5471\,A and C,
with FWZI(\halpha) $=$ 445\,\kms\ and 300\,\kms, respectively. With deeper echelle spectra, 
\citet{Chen:2002} measured FWZI(\halpha)\,$\simeq$\,620\,\kms\ for NGC~5471\,B.
Our measurements are in the range FWZI(\hbeta)\,$\simeq$\,400\,--\,1300\,\kms\
and FWZI(\oiii\lin5007)\,$\simeq$\,750\,--\,1300\,\kms. The largest values are found in NGC~5471\,C (\hbeta) and NW of the centre of NGC~5471\,A (\oiii\lin5007).

It is worth pointing out that the linear diameters of SNRs in M101 measured by \citet{Franchetti:2012} from {\em HST} images, and classified as due to core-collapse SNe, is in the range 20\,--\,80\,pc: at our seeing-limited resolution of about 30\,pc we can expect to barely resolve a typical SNR in this galaxy. High spatial resolution images by \citet{Chen:2002} show that the 
SNR in NGC~5471\,B, identified as a resolved \sii-bright shell, is approximately 2 arcsec in diameter, which is twice our seeing. 

Based on the finding by \citet{Sramek:1986} that both NGC~5471\,A and B have nonthermal radio emission, a signature of the presence of a SNR, 
CK86 suggested that all the three main knots in NGC~5471 (A, B and C) could host SNRs. The fact that these regions stand out so clearly in the maps of Fig.~\ref{fwzi_vs_broad}\,--\,\ref{FWZI} supports their suggestion.
FWZI(\oiii\lin5007) in NGC~5455\,A peaks at 750\,\kms, suggesting a similar interpretation, although we are unaware of the discovery of SNR signatures (nonthermal radio emission or indications of shocked gas from optical emission line ratios) in this object.
SN 1970G (whose location is shown in Fig.~\ref{n5455ha}) exploded to the NW of NGC~5455\,A.

In NGC~5471\,C \citet{Castaneda:1990} measured a very large FWZI(\halpha)\,$\simeq$\,3300\,\kms. We do not confirm such an extreme value, possibly due to the lower signal-to-noise ratio of our \hbeta\ line. Integrating the flux in a $2\times2$\,arcsec$^2$ region centred on this knot, in order to increase the signal-to-noise ratio,
we obtain FWZI(\hbeta)\,$\simeq$\,2100\,\kms, which is considerably larger than for any other knot ($\leq 750$\,\kms). We find FWZI(\oiii\lin5007)\,$\simeq$\,1450\,\kms. 
We thus confirm the presence of a remarkable low-intensity velocity extension of the Balmer lines in NGC~5471\,C.

We find it interesting that NGC~5471\,A, B and C, around which we detect large FWZIs, are 
three of the four most \halpha-luminous clusters in this complex  (\citealt{Garcia-Benito:2011}). 
However, the most \halpha-luminous cluster after knot A is knot A$^\prime$, which does not exhibit a particularly large FWZI or $\sigma$(broad). This suggests that the effect of something besides the energy output of `living' massive stars, namely supernova explosions, is required to generate a broad low-intensity profile.

\begin{figure*}
	\center
	\includegraphics[width=0.65\textwidth]{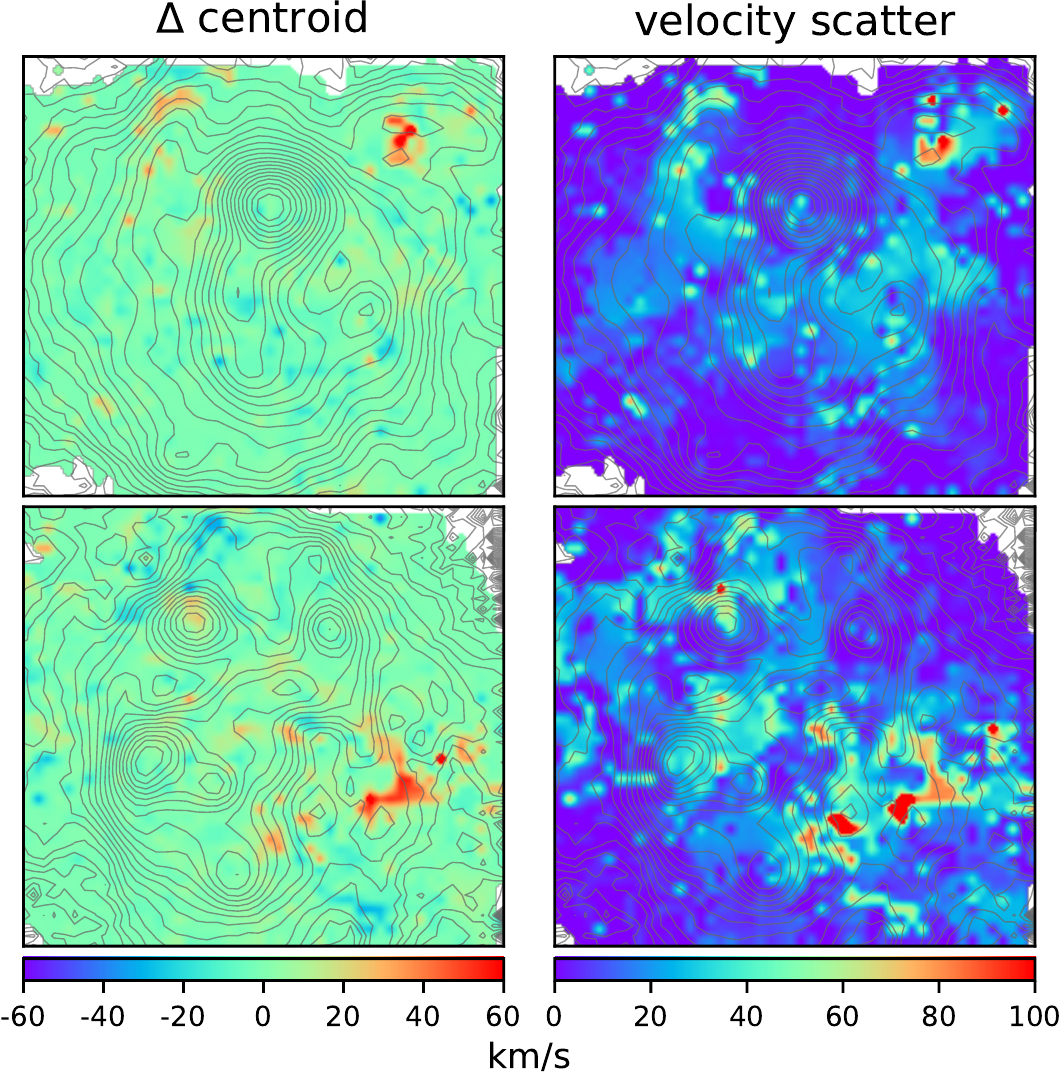}
	\caption{(Left) Maps of the velocity difference between the single Gaussian centroid and the broad component centroid. 
		(Right) Maps of the velocity scatter  
		between the narrow components revealed by the multiple Gaussian fits, measured by the rms dispersion of their central velocities.
		NGC~5455 is at the top and NGC~5471 is at the bottom. All maps refer to the \oiii\lin5007 line.
		\label{DeltaV}}
\end{figure*}

\subsubsection{Velocity structures and evidence for shells}
In the left panels of Fig.~\ref{DeltaV} we look at the velocity difference between the single Gaussian centroid (representative of the bright, narrow part of the line profile) and the broad component centroid (the midpoint of the FWZI range) for the \oiii\lin5007 line. Significant deviations (+60 \kms) between these quantities are localized in the `shell areas' NW of the centre in NGC~5455 and SW of the centre in NGC~5471, suggesting that at these locations the weak, broad profile component is decoupled from the stronger narrow components. In the rest of the field of view the difference is generally small and on average zero. 
These two areas match the locations of largest splitting of the composite line profiles, measured by the rms dispersion of the central velocities of the narrow components, shown in the map of the right panels of Fig.~\ref{DeltaV}. This figure provides similar information as, and is generally consistent with, the single Gaussian $\sigma$ maps shown in Fig.~\ref{isigma5455} and \ref{isigma5471}.

Since line splitting is expected in the presence of expanding gas shells or multiple moving filaments, the maps of the velocity scatter are consistent with the idea that the two areas defined earlier are populated by expanding shells and filaments moving at different velocities along the line of sight (see also our discussion of Figures~\ref{isigma5455}\,--\,\ref{isigma5471}), for which we measure an rms 
dispersion of the central velocities of the multiple narrow components
of up to $\sim$100\,\kms. Line splitting at the 20\,--\,40\,\kms\ level is widespread in our fields, a manifestation of the complex internal kinematics of these two giant \hii\ regions, as already demonstrated by the line profiles shown in Fig.~\ref{n5455littlegaussians}\,--\,\ref{n5471littlegaussians}.

\section{Global supersonic line widths}
We can now return briefly to the supersonic line widths exhibited by the global profiles presented in Sect.~\ref{Sec:globalprofiles}, and which is typical of giant \hii\ regions. 
For the best-studied case, 30~Dor, \citet{Chu:1994} and \citet{Melnick:1999}
demonstrated that the integrated profile breaks up into discrete components of subsonic widths. 
\citet{Melnick:2019} have shown that the supersonic width of 30~Dor is in fact due to macroscopic gas elements
moving inside the gravitational potential of the region. 
In addition, the presence of stellar winds plays a considerable role in the overall dynamics. 

\citet{Melnick:2019} tested this conclusion, by showing that the integrated line width is comparable to the rms dispersion of the radial velocity measured across the face of the nebula. This procedure cannot be used in our case: the velocity dispersion is `diluted' by the $\sim$140$\times$ coarser physical spatial resolution, down to 
$5-8$\,\kms. We have instead performed a consistency test, by calculating the intensity-weighted rms dispersion obtained from the velocity decomposition of the line profiles (the single-spaxel results are shown in Fig.~\ref{DeltaV}).
Although we cannot expect perfect agreement with the width of the integrated line profiles, due to the unavoidable difficulties intrinsic to multiple Gaussian fits, we should find a similar value. We limit the comparison to the \oiii\lin5007 lines, because of the higher signal-to-noise ratio, but above all because of the much smaller smearing from thermal broadening. For NGC~5455 we find a weighted rms dispersion of 
15.7\,\kms, and for NGC~5471 a value of 22.2\,\kms. These results compare favorably with the integrated line widths (see Table~\ref{tab:comparison}), particularly in the case of NGC~5471.

We conclude that our data provide evidence that the major contributor to the supersonic turbulence measured in giant \hii\ regions (excluding the underlying components) is the motion of discrete macroscopic cloudlets, in agreement with \citet{Melnick:2019}.

\medskip
The stellar masses of our target \hii\ regions can be obtained from their narrow-band luminosities from

\begin{equation}
\log(M^*/M_\odot) =  \log[L(H\beta)] - 33.15.
\end{equation}

\noindent
This equation was obtained from instantaneous burst models calculated with Starburst99 (\citealt{Leitherer:1999}), integrating a Salpeter initial mass function over the range 0.2\,--\,100\,\msun.
From the \hbeta\ luminosities published by \citet{Fernandez-Arenas:2018} we derive very similar masses for 
NGC~5455 and NGC~5471, $\log(M^*/M_\odot) \sim 6.7-6.8$.

It is instructive to compare these photometric masses with estimates of the dynamical masses, assuming that their global gas motions are due to gravity.
From the virial theorem the velocity dispersion is related to the mass as follows:

\begin{equation}
\sigma = \left( \frac{GM_d}{\eta \reffmath} \right)^{0.5}
\end{equation}

\noindent
where $\reffmath$ is the effective radius enclosing half of the mass,
and $\eta$ is on the order unity.
Taking a stellar mass of $6\times10^6$\,\msun, derived from the ionizing flux, we obtain $\sigma\simeq 21$\,\kms, close to the observed global values (see Table~\ref{tab:comparison}), if we assume $\reffmath = 60$\,pc. Although we cannot measure $\reffmath$ directly, this value appears plausible.

\section{The supernova remnant in NGC~5471\,B}\label{Sec:knotB}

NGC~5471\,B hosts a SNR, as first shown by \citet{Skillman:1985}.
We detect a bright and broad line component associated with it across a region of $\sim$90\,pc in diameter. 

\begin{figure}
	\center
	\includegraphics[width=0.9\columnwidth]{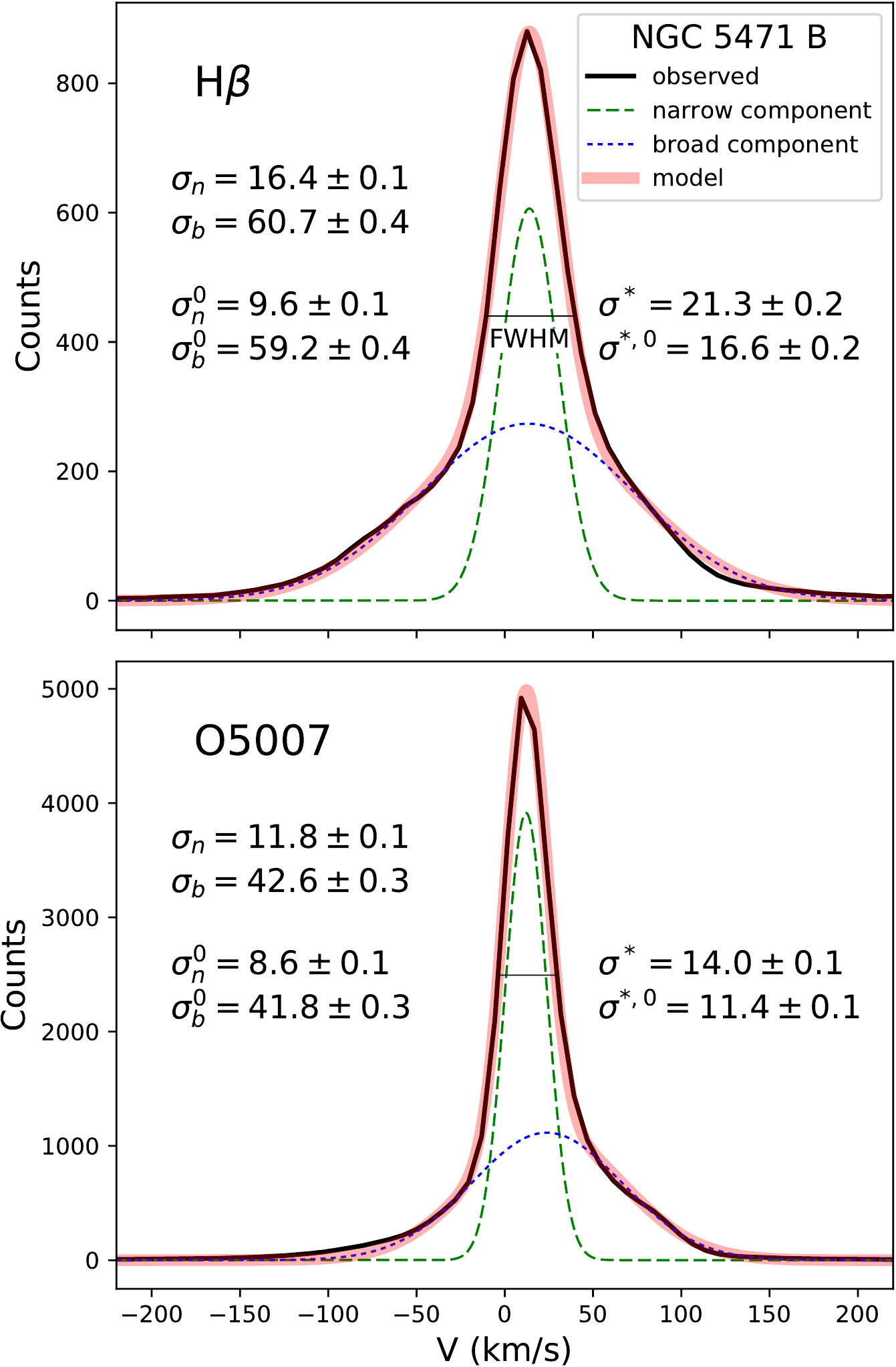}
	\caption{Integrated \hbeta\ (top) and \oiii\lin5007 (bottom) line profiles of NGC~5471\,B (black profile). The decomposition into a narrow (green dashed line) and a broad (blue dotted line) Gaussian components results in the model shown in red. The measured widths $\sigma$ are reported, in \kms, for both the narrow ($n$) and the broad ($b$) components.	Widths corrected for thermal and instrumental effects (plus fine structure for \hbeta) are indicated as $\sigma^0$. 
		A `pseudo standard deviation' $\sigma^*$ is measured from the FWHM of the model (indicated), with $\sigma^{*,0}$ its corrected value.
		\label{FigKnotB}}
\end{figure}

The \hbeta\ and \oiii\lin5007 line profiles, integrated over a $2\times2$\,arcsec$^2$ region centred on the \hbeta\ peak, are displayed in Fig.~\ref{FigKnotB}. We have decomposed each profile using a two-component Gaussian fit, which works reasonably well in this restricted area, in order to focus on the two main features: the narrow central component, attributed to the host \hii\ region (dashed line) and the high-intensity broad component, attributed to the SNR (dotted line)  (we are therefore not fitting the even broader low-intensity wings that were analysed in Sect.~\ref{Sec:broadcom}). 
In the case of \hbeta\ we find a narrow component line width, corrected for thermal, instrumental and fine structure effects, of $\sigma_n^0$(\hbeta) = $9.6 \pm 0.1$\,\kms, while for the broader component we measure $\sigma_b^0$(\hbeta) = $59.2 \pm 0.4$\,\kms.
CK86 reported an uncorrected value $\sigma_b$(\halpha) = $61.6$\,\kms, virtually equal to our $\sigma_b$(\hbeta) = $60.7 \pm 0.4$\,\kms. \citet{Chen:2002} measured $\sigma_n$(\halpha) = $17.4 \pm 0.8$\,\kms\ (compared to our $16.4 \pm 0.1$\,\kms)
and $\sigma_b$(\halpha) = $62.8 \pm 2.1$\,\kms. There is therefore full consistency between these independent measurements.

We notice several differences between the \hbeta\ and \oiii\ profile decompositions.
For \oiii\lin5007 both components are narrower than in the case of \hbeta: $\sigma_n^0$(\oiii) = $8.6 \pm 0.1$\,\kms\ and 
$\sigma_b^0$(\oiii) = $41.8 \pm 0.3$\,\kms. The flux of the broad component relative to that of the narrow component is also smaller: $f_b/f_n$(\oiii) = 1.03 \vs\ $f_b/f_n$(\hbeta) = 1.67. Moreover, while for \hbeta\ the broad component is well aligned with the velocity of the narrow component, for \oiii\ the broad component is shifted to higher velocities (by $\sim$12\,\kms).
We point out that for both lines the integrated narrow component has a slightly subsonic line width.

\begin{figure*}
	\center
	\includegraphics[width=0.8\textwidth]{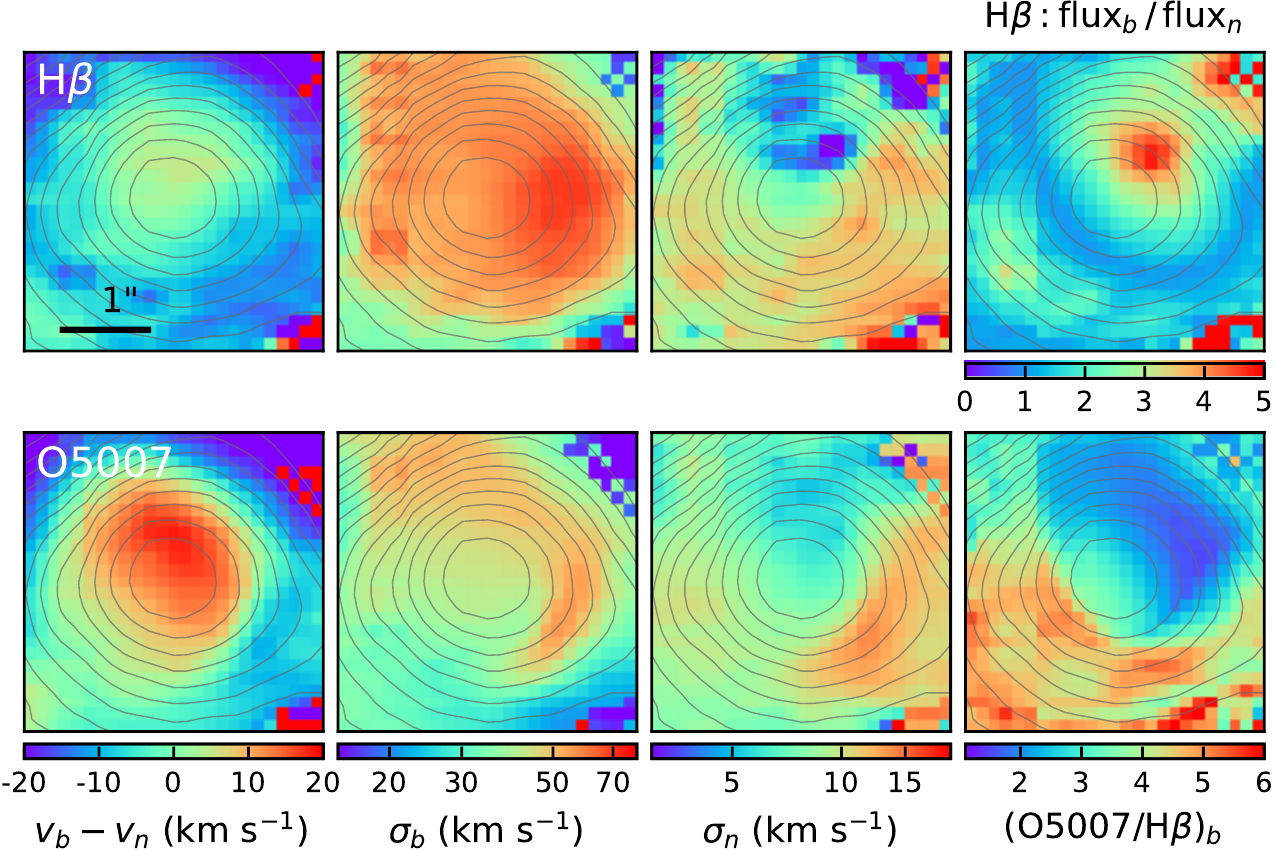}
	\caption{Kinematic and excitation maps of NGC~5471\,B derived from two-component fits to the \hbeta\ (top row) and \oiii\lin5007 (bottom rows) lines. The quantities shown in the first three columns are the velocity offset between the broad and narrow components ($v_b - v_n$), and the line widths of the broad ($\sigma_b$) and narrow ($\sigma_n$) components. In the last column we show the relative \hbeta\ flux between the two components (top) and the \oiii/\hbeta\ line ratio for the broad component (bottom).
		\label{knotBmaps}}
\end{figure*}


In the  case of NGC~5471\,B a single Gaussian fails miserably to fit the line profile, due to the presence of the intense broad component. In order to describe the overall line width
we adopt the FWHM measured from the model profile (indicated in Fig.~\ref{FigKnotB}) and define a `pseudo standard deviation',  $\sigma^*\,  \equiv \, $FWHM/2.355. For lines that are approximately following a Gaussian profile, $\sigma \simeq \sigma^*$, but in the presence of significant deviations from the Gaussian line shape $\sigma^*$ is a better representation of the line width.
We can see from the numerical results reported in Fig.~\ref{FigKnotB} that at half peak intensity the \hbeta\ line is 5.2\,\kms\ wider than \oiii, which results from the combined contribution of the larger values of both $\sigma_n$ and $\sigma_b$. 

Deeper insight into the kinematics and excitation properties of NGC~5471\,B can be gained from the maps shown in Fig.~\ref{knotBmaps}. These refer to a $3.5\times3.5$\,arcsec$^2$ section of the full-resolution data cube, centred on the \hbeta\ emission peak. The panels in this figure display maps of different quantities measured from 
a two-component fit to \hbeta\ (top row) and \oiii\lin5007 (bottom row). The first column of Fig.~\ref{knotBmaps} shows how the velocity offset between narrow and broad components  
is negligible for \hbeta, but peaks around 20\,\kms\ for \oiii\ -- a behaviour already established from the integrated line profiles of Fig.~\ref{FigKnotB}. The width of the broad components peaks to the west of the \hbeta\ emission peak (second column), but is significantly larger for \hbeta. Differences such as these are understood as a consequence of the fact
that the two emission lines are formed at different temperatures in 
the recombination flow associated with a radiative supernova shock (\citealt{Lasker:1977}), and therefore they refer to different gas filaments.
In the third column we show maps of the width of the narrow component $\sigma_n$. They are
completely different from the $\sigma_b$ maps, showing similarity between \hbeta\ and \oiii, as well as subsonic line widths in the central, brightest parts of NGC~5471\,B. 

In the top panel of the final column we show a map of the flux ratio between the broad and the narrow \hbeta\ components. The ratio peaks in the NW sector. The bottom panel of the same column shows that in the same NW sector the gas excitation of the broad component, measured by the \oiii/\hbeta\ line ratio, reaches its minimum value. In Sect.~\ref{Sec:global} we explained such a low \oiii/\hbeta\ ratio with the presence of an interstellar shock at low metallicity. With the help of the {\em HST}-based study by \citet{Chen:2002} 
we identify the low-excitation sector with the shell revealed by their \sii/\halpha\ line ratio map. 
The western edge of this $\sim$2 arcsec-wide shell, where the \sii/\halpha\ is higher, corresponds approximately to the position where we measure the largest $\sigma_b$. 
The top right panel of 
Fig.~\ref{knotBmaps} then shows that the strength of the broad component relative to the narrow component peaks near the centre of the shell.

\section{\hbeta\ \vs\ \oiii\ line widths}\label{Sec:o3vshbeta}
\begin{figure*}
	\centering
	\begin{minipage}[t]{.38\textwidth}
		\includegraphics[width=\columnwidth]{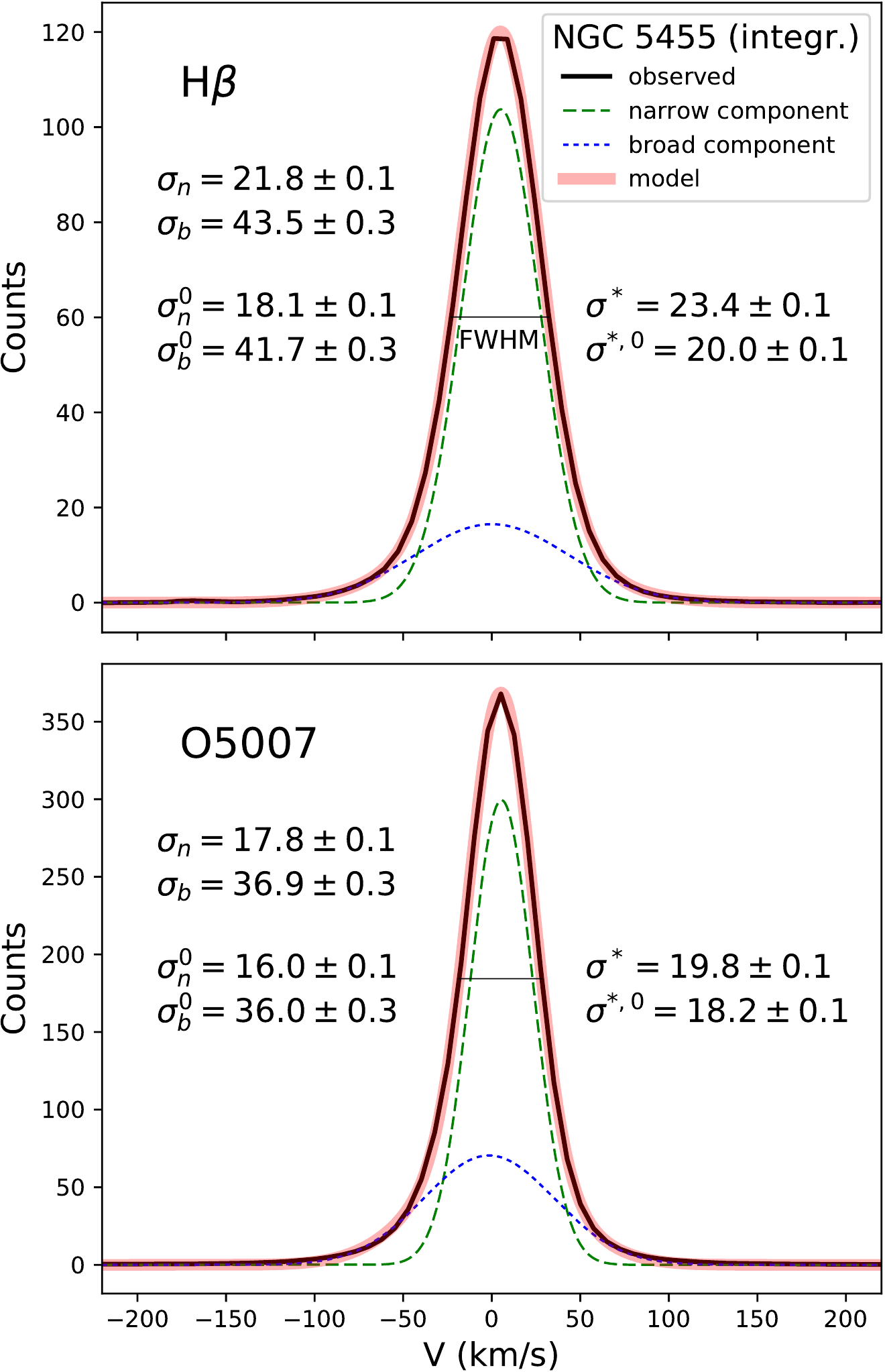}
	\end{minipage}\qquad
	\begin{minipage}[t]{.38\textwidth}
		\includegraphics[width=\columnwidth]{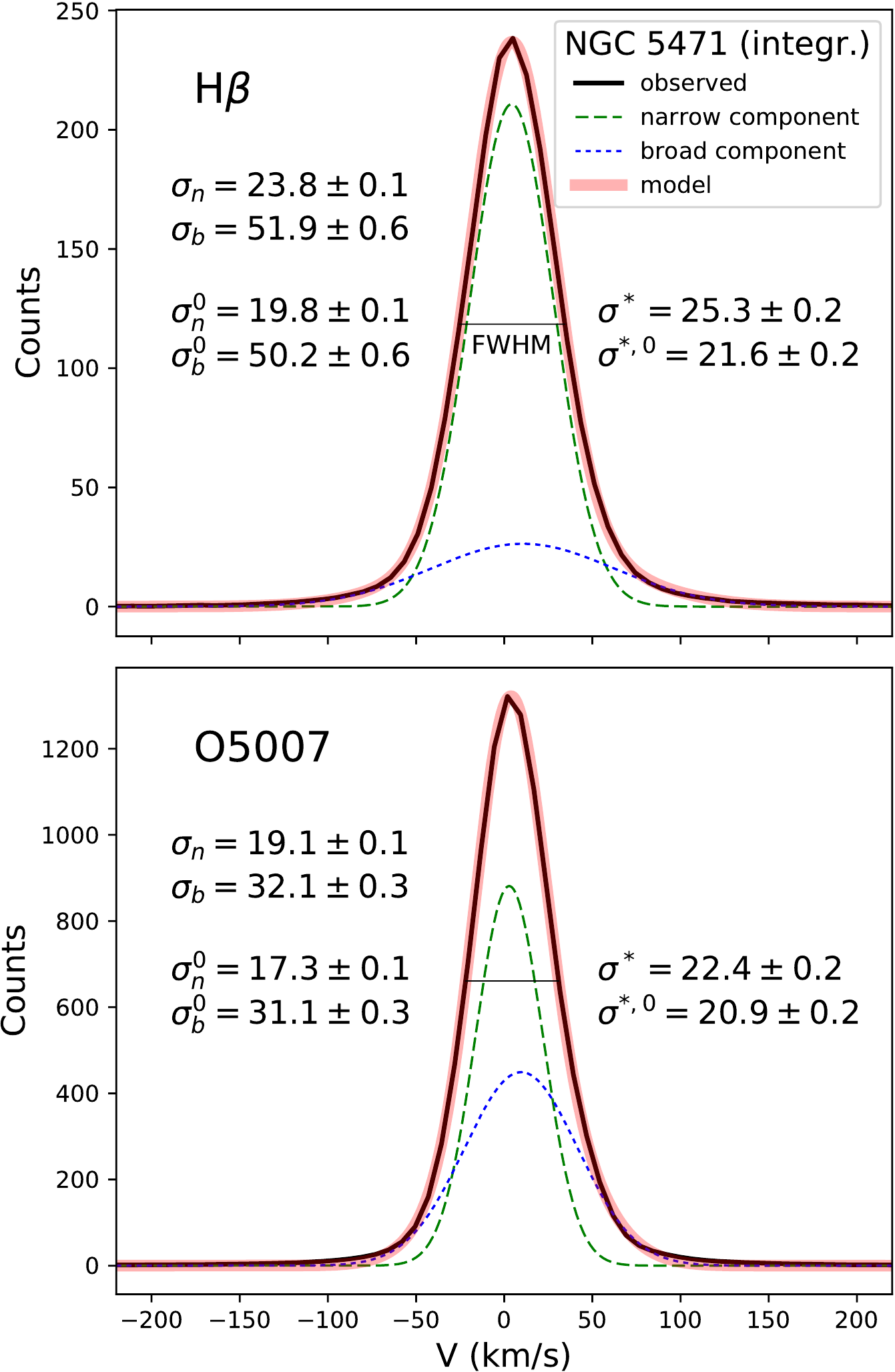}
	\end{minipage}
	\caption{Two-Gaussian decomposition of the integrated \hbeta\ and \oiii\lin5007 line profiles for NGC~5455 (left) and NGC~5471 (right). See Fig.~\ref{FigKnotB} for an explanation of the different line types and  quantities reported.}\label{IntegratedBoth}
\end{figure*}

\citet{Hippelein:1986} noted that the integrated line widths of extragalactic giant \hii\ regions, corrected for thermal broadening, differ systematically between \hi\ and \oiii\ lines, finding $\sigma(H\alpha)-\sigma$(\oiii\lin5007) $\simeq 2$\,\kms. A comparable result was obtained for \hii\ galaxies by \citet{Bordalo:2011}. 
Our integrated measurements for NGC~5455 and NGC~5471, presented in Table~\ref{tab:comparison}, are in agreement with these findings. 
Moreover, in a fashion similar to our profile decomposition for NGC~5471\,B, presented in Sect.~\ref{Sec:knotB} and shown in Fig.~\ref{FigKnotB}, we  carried out a two-Gaussian fit to the integrated 
\hbeta\ and \oiii\lin 5007 line profiles of both NGC~5455 and NGC~5471. The results are presented in Fig.~\ref{IntegratedBoth}. It can be seen that the line width difference between \hbeta\ and \oiii\lin5007
is confirmed for both the (corrected) narrow components ($\sigma_n^0$) and the composite model width ($\sigma^{*,0}$), albeit with somewhat differing magnitudes.

We inspected the line width discrepancy between \hbeta\ and \oiii\ in the brightest star-forming knots of our two targets (NGC~5471\,A, A$^\prime$, B, C, E and NGC~5455\,A), defining $2\times2$ arcsec$^2$ regions centred at their positions. We decomposed their line profiles, following the method outlined
in Sect.~\ref{Sec:MultiGauss}, and calculated the FWHM of the model profile. As explained in Sect.~\ref{Sec:knotB} we define $\sigma^*\,  \equiv \, $FWHM/2.355, obtaining a mean, corrected value difference $\sigma^{*,0}(H\beta)-\sigma^{*,0}$(\oiii\lin5007) = $1.7 \pm 0.8$\,\kms\ (we excluded NGC~5471\,B, for which the difference is considerably larger than the rest: 5.2\,\kms\ -- see Sect.~\ref{Sec:knotB}). The result is virtually unchanged using single Gaussian fits: 
$\sigma(H\beta)-\sigma$(\oiii\lin5007) = $1.5 \pm 0.7$\,\kms, with the distinction that the profile decomposition reveals the separate contribution of narrow and broad components to the overall profile.
In conclusion, the tests we carried out confirm a line width offset of about 1\,--\,2\,\kms\  between the \hbeta\ and \oiii\ line profiles, both globally and for individual star forming cores.

\begin{figure*}
	\center
	\includegraphics[width=0.83\textwidth]{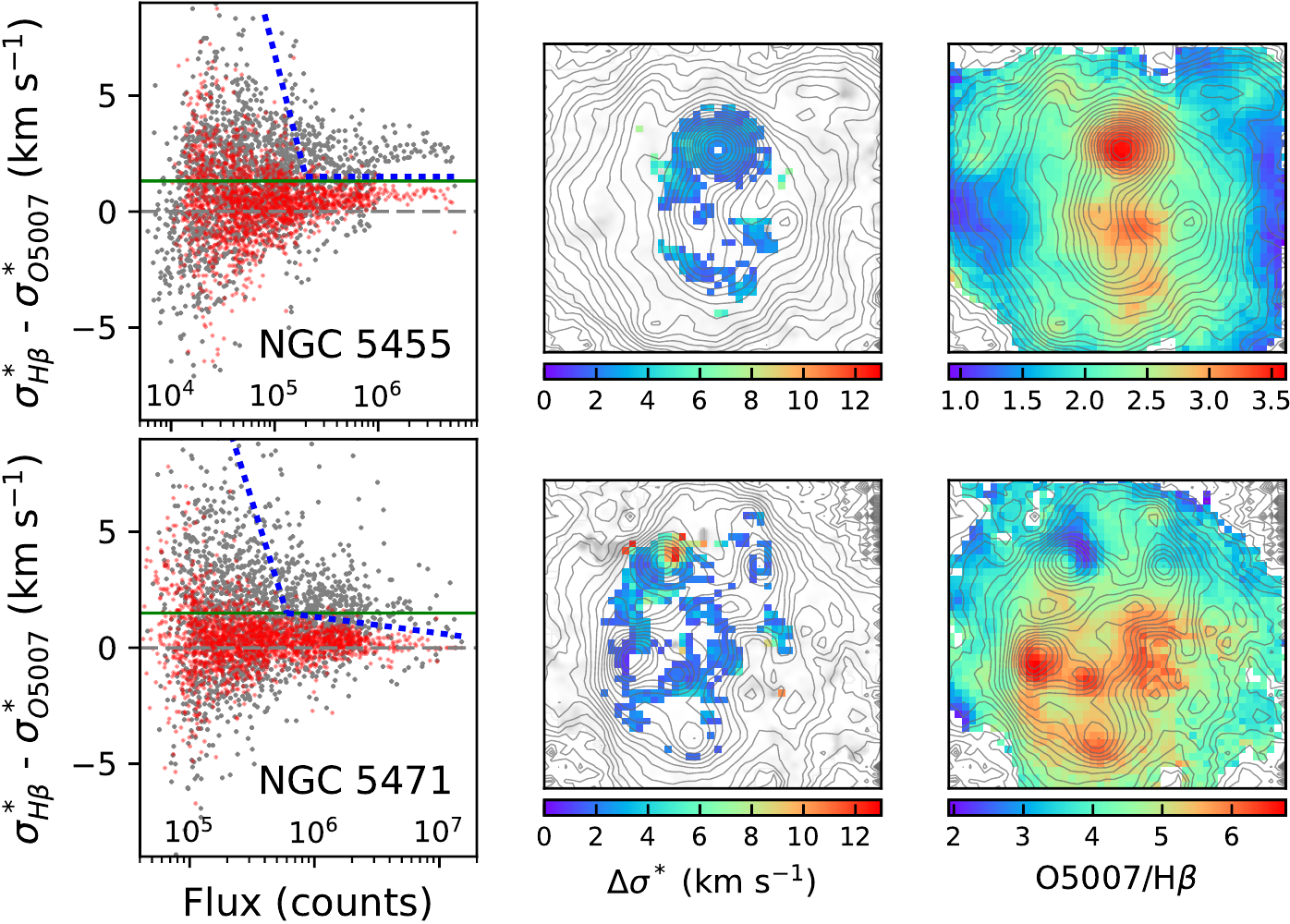}
	\caption{(Left) Line width difference $\Delta\sigma^* = \sigma^*(H\beta)-\sigma^*$(\oiii\lin5007) as a function of the \oiii\lin5007 flux measured from multiple Gaussian line fits (grey dots). Red dots represent the width difference measured between \oiii\ \lin4959 and \lin5007 lines, in order to estimate the $\Delta\sigma^*$ uncertainties.
		The horizontal dashed line is drawn at $\Delta\sigma=0$, while the mean $\Delta\sigma^*$ difference is shown by the continuous green line. The blue dotted lines isolate the spaxels for which $\Delta\sigma^*$ is considered to be significant.
		NGC~5455 is at the top, NGC~5471 at the bottom.
		(Center) Maps of $\Delta\sigma^*$ for the spaxels isolated in the left panels. (Right) Maps of the \oiii\lin5007/\hbeta\ line flux ratio.
		\label{deltasigma}}
\end{figure*}

We examined how the \hbeta\ and \oiii\ line width difference varies with position in NGC~5455 and NGC~5471, by taking the results of our multiple Gaussian fits (Sect.~\ref{Sec:MultiGauss}), and considering the $\sigma^*$ widths of the model profiles. The grey dots in the left panels of Fig.~\ref{deltasigma} represent the difference
$\Delta\sigma^* = \sigma^*(H\beta)-\sigma^*$(\oiii\lin5007) measured in each spaxel. 
The increasing spread with decreasing line flux is largely due to random uncertainties in our measurements. In order to 
assess the magnitude of these uncertainties, we represent with red dots the $\sigma^*$ difference obtained using the \oiii\ lines at \lin4959 and \lin5007. These two lines should yield the same width, so that measurement uncertainties are now (approximately) represented in the plot. The comparison between grey and red dots indicates that 
only positive $\Delta\sigma^*$ values for the brightest spaxels should be deemed as significant. We conservatively isolated these spaxels to be those contained in the area of the plot above the dotted line. Their spatial location and 
$\Delta\sigma^*$ values are shown in the central panels of Fig.~\ref{deltasigma}. Most points correspond 
to $\Delta\sigma^* \simeq 2-3$\,\kms, with the notable exception of NGC~5471\,B, where the width difference increases to 13\,\kms\ -- as shown in Sect.~\ref{Sec:knotB} this is related to the strong contribution from the SNR to the line profile. 
The right panels display the gas excitation map, using the \oiii\lin5007/\hbeta\ line ratio. We fail to discern obvious correlations between $\Delta\sigma^*$  and gas excitation (except again for the case of NGC~5471\,B, where the excitation reaches a minimum -- see Sect.~\ref{Sec:knotB}), as well as differences  between star-forming cores and diffuse gas. Admittedly, the fractional coverage of the
spaxels that survived our selection in the central panel of Fig.~\ref{deltasigma} is small, but to our knowledge this is the first time that the width discrepancy between hydrogen and metal lines is examined on a spatially-resolved basis.

The explanation for the systematic line width difference between recombination and collisionally excited lines of photoionized nebulae continues to elude us. 
However, we find it remarkable that the line width difference 
is quantitatively comparable ($\sim$2\,\kms) among giant \hii\ regions
and the more luminous ensembles of giant ionized nebulae in \hii\ galaxies, and between giant \hii\ regions and objects like the Orion nebula (\citealt{ODell:2003}) that are more than three orders of magnitude less luminous. 
These similarities suggest a common origin.
The disparity of kinematical conditions between small and giant \hii\ regions is such that 
smearing of fine details of non-Gaussian line profiles by thermal broadening (\citealt{Garcia-Diaz:2008}) appears as a less likely explanation than an interpretation based on physical attributes of the nebular gas (\citealt{ODell:2017}). This conclusion is supported by the fact that, at least in the Orion nebula, 
\hi\ and \hei\ recombination lines have virtually the same width (\citealt{ODell:2017}), while
a significant reduction of line smearing due to thermal broadening (approximately a factor of two according to Eq.~\ref{Eq:thermal}) should be observed for \hei\ lines instead.
Work is under way to address these issues in a future publication concerning a sample of giant extragalactic \hii\ regions.

\section{Summary}

We have studied the kinematics of the giant \hii\ regions NGC~5455 and NGC~5471, located in the galaxy M101, using integral field data obtained with the Keck Cosmic Web Imager. We have fitted the profiles of the \hbeta\ and \oiii\lin5007 lines using both single Gaussian curves and multiple Gaussian models. While the former technique cannot capture all the detailed information that the multiple profile approach can provide, the interpretation of the two types of fits has points in common that offer consistent results, but some important divergences emerge. Our main findings are listed below:\\[-3mm]

\begin{enumerate}

\item \noindent 
both the $I-\sigma$ diagram (from single Gaussian fits) and the multiple profile analysis provide
kinematic evidence for the presence of several expanding shells and moving filaments. 
\\[-5pt]

\item \noindent
decomposing the line profiles using multiple Gaussians yields a large fraction of components with subsonic line widths. The single Gaussian fits instead provide  only supersonic line widths -- this outcome underlines the inadequacy of fitting single Gaussian curves to the line profiles of giant \hii\ regions. We cannot exclude that at higher spectral and spatial resolution multiple Gaussian fits would split all `narrow' supersonic components into subsonic ones.
\\[-5pt]

\item \noindent
an underlying broad component characterized by velocity widths $\sigma\simeq 30-50$\,\kms\ -- possibly resulting from stellar wind interactions with cold gas -- is commonly found in the bright parts of our two targets, extending across hundreds of pc. This component can be described by the full width at zero intensity of the line profiles, a quantity that is independent of the accuracy of the multiple line fitting.
\\[-5pt]

\item \noindent the supersonic turbulence inferred from the global profiles is consistent with the combined contribution of individual gas cloudlets, and the dispersion of their velocities.
\\[-5pt]

\item \noindent
for three bright star-forming knots in NGC~5471 (A, B and C) the large velocity spread (750\,--\,1300\,\kms\ in the case of \oiii)
of the low-intensity component supports previous suggestions that all these regions host supernova remnants.
\\[-5pt]

\item \noindent
known and suspected supernova remnants in NGC~5471 stand out thanks to their low \oiii/\hbeta\ line ratios relative to the surrounding gas -- we attribute this to the low metallicity of the region.
\\[-5pt]

\item \noindent
the known supernova remnant in NGC~5471\,B, barely spatially resolved in our data, displays remarkable differences between \hi\ and \oiii\ emission lines.
\\[-5pt]

\item \noindent
we confirm the presence of a global, as well as a localized, systematic offset between the widths of the \hbeta\ recombination line and the \oiii\lin5007 collisionally excited line, which is independent of the gas excitation. We argue that its origin lies in the physics of the line-emitting gas rather than the larger thermal broadening of the Balmer lines.

\end{enumerate}



\section*{Acknowledgments}
The data presented herein were obtained using KWCI at the W. M. Keck Observatory, which is operated as a scientific partnership among the California Institute of Technology, the University of California and the National Aeronautics and Space Administration. The Observatory was made possible by the generous financial support of the W. M. Keck Foundation.
The authors wish to recognize and acknowledge the very significant cultural role and reverence that the summit of Maunakea has always had within the indigenous Hawaiian community.  We are most fortunate to have the opportunity to conduct observations from this mountain.
We thank the referee Casiana Mu\~noz-Tu\~n\'on for her comments, that helped us to improve our manuscript.

\bibliographystyle{mnras}
\bibliography{/Users/fabio/PDF-Papers/Papers}

\bsp	
\label{lastpage}
\end{document}